\documentclass[aps,pra,twocolumn,letterpaper,superscriptaddress]{revtex4-2}

\usepackage{amsmath}
\usepackage{amssymb}
\usepackage{amsfonts}
\usepackage{bm}
\usepackage{mathrsfs}
\usepackage{tensor}
\usepackage{dsfont}
\usepackage{esint}
\usepackage{comment}
\usepackage{graphicx}

\usepackage{multirow}

\usepackage[usenames,dvipsnames]{xcolor}
\usepackage[hyperindex,pdftex, breaklinks,colorlinks = true,linkcolor = blue,urlcolor=blue,citecolor=blue]{hyperref}

\usepackage{physics}
\usepackage[section]{placeins}

\begin{document}
	
	\title{Chern insulator optics}

	\author{Jason G. Kattan}
	\email{jason.kattan@mcgill.ca}
    \affiliation{Department of Physics, McGill University, Montréal, Québec H3A 2T8, Canada}
	\affiliation{Department of Physics, University of Toronto, Toronto, Ontario M5S 1A7, Canada}
 
	\author{J. E. Sipe}
	\email{john.sipe@utoronto.ca}
	\affiliation{Department of Physics, University of Toronto, Toronto, Ontario M5S 1A7, Canada}
	
	\date{\today}
	
	\begin{abstract}
        Chern insulators have optical properties that are strikingly different than ordinary chiral media. Unlike the natural optical activity of the latter, which is a finite-wavelength effect arising from spatial dispersion, the optical activity of a Chern insulator is already present in the long-wavelength limit, and exhibits a striking directional asymmetry in the propagation of circularly polarized light. We study this “unnatural” optical activity using a recently developed field theory for Chern insulators. For bulk Chern insulators, we derive the normal-incidence transmission and reflection coefficients of multilayer structures, identify the associated Faraday and Kerr angles, and show that these optical rotations provide a direct optical signature of topology at low frequencies. We also show that an interface between bulk Chern insulators can act as a kind of mirror, completely reflecting light at normal incidence when the Chern invariants have opposite signs. Moving to a more realistic model of a two-dimensional current sheet, we study the Faraday and Kerr effects in a thin film of $\mathrm{Mn}\mathrm{Bi}_2\mathrm{Te}_4$ in its Chern insulator phase. The predicted rotations are many orders of magnitude larger than those of optically active media of comparable thickness that are topologically trivial.
    \end{abstract}
	
	\maketitle

%%%%%%%%%%%%%%%%%%%%%%%%%%%%%%%%%%%%%%%%%%

\section{Introduction}\label{Sec:Introduction}

When light passes through a material, it carries more than just an intensity and phase. It also carries a polarization, and in many materials this polarization can change in a way that reveals the internal structure of the medium. For transparent media, in the simplest case there is a single index of refraction, and light of a given frequency will propagate with a phase that is independent of its polarization \cite{BornWolf1999}. Birefringent crystals can distinguish between different linear polarizations, while in optically active media left-handed and right-handed circularly polarized beams will propagate with different phase velocities \cite{Nye1985}. The latter effect is familiar in crystals such as $\alpha$-quartz \cite{Lekner1996}; there, the optical activity is associated with a term in the response that is first order in the wavevector of the light, and for a beam propagating in one direction with a given circular polarization, the beam propagating in the opposite direction with the same circular polarization propagates with the same refractive index \cite{Landau1984}. This kind of ``natural" optical activity is a finite-wavelength correction to local optical response, and is the result of spatial dispersion; it disappears in the long-wavelength limit. But in a Chern insulator the situation is strikingly different.

Chern insulators are one class of the broader family of topological insulators that have attracted much interest in the condensed matter community in recent years \cite{Hasan2010,Qi2011}. They are band insulators with spontaneously broken time-reversal symmetry, often as a result of some kind of long-range magnetic order in the crystal, and are characterized by one or more integer-valued \textit{Chern numbers.} Their values are identified with the topological properties of the occupied valence bands in the crystal's band structure \cite{Thouless1982,Haldane1988}. The transport properties of Chern insulators are well-understood \cite{Chang2023}. When a DC electric field is applied, a transverse current is generated that is topologically quantized, depending only on the Chern numbers and nothing else; this is the quantum anomalous Hall effect \cite{Yu2010}. But the way in which a Chern insulator responds to light is more subtle. 

The optical activity that arises in these Chern insulators is qualitatively different than the natural optical activity that occurs in $\alpha$-quartz, or in solutions of chiral molecules \cite{Barron2004}. In the ``unnatural" optical activity of a Chern insulator, the difference between the refractive indices for left-handed and right-handed circularly polarized light arises at zeroth order in the wavevector of light and, for a beam propagating in one direction with a given circular polarization, the beam propagating in the opposite direction with the same circular polarization propagates with the \textit{other} refractive index. Since the optical activity in a Chern insulator arises in the long-wavelength limit, the effect can be much larger than in ``naturally" optically active materials, and manifests itself in very large values of the Faraday and Kerr angles that characterize optical rotation of the transmitted and reflected fields, respectively \cite{Tse2010,Tse2011}. Indeed, even in the limit of a nanometer-thickness film, the Faraday angle for the transmitted light is on the order of a degree; this is a massive rotation angle compared to the optically active yet topologically trivial $\alpha$-quartz, for example, which under the same conditions is on the order of $10^{-5}$ degrees \cite{Lekner1996}. 

There are several materials that have been found to support a Chern insulator phase, the first examples of which were thin films of magnetically doped $\mathrm{Bi}_2\mathrm{Te}_3$ and $\mathrm{Sb}_2\mathrm{Te}_3$ \cite{Chang2013,Chang2015}; magnetic doping leads to the necessary breaking of time-reversal symmetry. More recently, it has been found that thin films of the intrinsic magnetic material $\mathrm{Mn}\mathrm{Bi}_2\mathrm{Te}_4$ that consist of an odd number of septuple layers become Chern insulators at sufficiently low temperatures \cite{Zhang2019,Otrokov2019a,Otrokov2019b,Liu2020}. Following advances in their fabrication, a number of novel experiments on the optical properties of these materials have been performed \cite{Tomarchio2022,Kopf2020,Kopf2022}. There have also been a few theoretical studies concerning the optics of Chern insulators \cite{Qi2008,Maciejko2010,Tse2010,Tse2011,Zhu2023,Ghosh2024}, which primarily focus on the transmission and reflection of light across thin films of materials like $\mathrm{Mn}\mathrm{Bi}_2\mathrm{Te}_4$, often based on calculations employing tight-binding models with effective Hamiltonians \cite{Lei2023}. 

We follow a different strategy here, based on a generalization of the ``modern theories of polarization and magnetization" \cite{Resta2011,Vanderbilt2018} to treat the response of crystalline media to time-dependent fields \cite{Mahon2019,Mahon2023}. Adopting a microscopic model involving cell-periodic effective magnetic fields that are responsible for the breaking of time-reversal symmetry \cite{Kattan2026}, we can address the optical properties of a Chern insulator using a long-wavelength conductivity tensor that was derived within this formalism \cite{Kattan2025}. This conductivity tensor can be written as a sum of two contributions. The first involves the Chern invariant but has no frequency dependence, and describes the quantum anomalous Hall effect; the second can be written in terms of a susceptibility tensor that accounts for the finite-frequency part of the response, including absorption. Both contributions lead to optical activity. At 
sufficiently low frequencies the topological Hall term involving the Chern invariant dominates, while at higher frequencies closer to the bandgap the susceptibility, which has an asymmetric index structure that is a direct consequence of the broken time-reversal symmetry, is dominant. In this sense, our conductivity tensor identifies the optical properties of the Chern insulator. The optical response of an isolated Chern insulator, or of a Chern insulator in any multilayer structure, can then be addressed using the standard approach of thin-film optics \cite{Jackson1999}. 

In this paper we address the novel optical properties of Chern insulators, focusing on frequencies below the band gap where there is no absorption. We begin in Sec.~\ref{Sec:Cherninsulators} with a brief summary of a field theory for Chern insulators introduced in recent work \cite{Kattan2026}, in terms of which the conductivity tensor was derived \cite{Kattan2025}. From this we identify an effective dielectric tensor in Sec.~\ref{Sec:Noveloptics} and, considering the propagation of light in such a medium, we characterize the long-wavelength optical activity that results and contrast its behavior with the more familiar natural optical activity that arises from spatial dispersion. In Sec.~\ref{Sec:Bulkoptics} we calculate the transmission and reflection coefficients for light at normal incidence on a multilayer structure containing a three-dimensional Chern insulator; we defer the treatment of non-normal incidence to a later 
communication. We can use these coefficients to calculate the Faraday and Kerr angles that characterize the rotation of the polarization of light as it passes through an optically active medium. We also show that these quantities are ``topologically quantized" at sufficiently long wavelengths, and thus constitute a direct optical probe of topology in bulk Chern insulators, without having to resort to the confusing notion of ``chiral edge modes" in the bulk-boundary correspondence. And in Sec.~\ref{Sec:Chernmirror} we consider light incident on an interface separating two bulk Chern insulators, and we point out the novel features of these ``Chern mirrors."

Admittedly, three-dimensional bulk Chern insulators have yet to be realized in the laboratory. Even the septuple layers of $\mathrm{Mn}\mathrm{Bi}_2\mathrm{Te}_4$ that have been produced are of a thickness much less than the wavelength of light. For these materials -- whether they stand alone or are part of a larger multilayer structure -- a better model is that of a current sheet describing the response of a Chern insulator in two dimensions. This we present in Sec.~\ref{Sec:Sheetoptics}, and we derive expressions for the optical response at normal incidence of a multilayer structure containing a Chern insulator, comparing these expressions with our earlier results for a bulk Chern insulator. We summarize our results in Sec.~\ref{Sec:Discussion}, and indicate directions for future research. Details of the calculations and some of the terminology can be found in the Appendices.

\section{Chern insulators}\label{Sec:Cherninsulators}

We begin by summarizing a recently developed field theory for Chern insulators \cite{Kattan2026}. The long-wavelength response of a Chern insulator was studied in the context of this field theory in other work \cite{Kattan2025}; we use the results, which are summarized below, in our study of Chern insulator optics. The calculations are performed within the independent-particle approximation, and thus are restricted to treat integer Chern insulators that can be described at a Hartree level.

\subsection{Hamiltonian theory}\label{Sec:Hamiltoniantheory}

We consider a Chern insulator with a lattice $\Gamma$ that is extended in $d$ spatial dimensions; the insulator is initially occupying its electronic ground state at zero temperature. We make the frozen-ion approximation, where the ions in the crystal are taken to be in a fixed configuration in space, and so the electrostatic interactions between the electrons and the ions in the crystal are described by a potential energy $\mathrm{V}_{\Gamma}(\bm{x})$ having the periodicity of the crystal's lattice, with $\mathrm{V}_{\Gamma}(\bm{x} + \bm{R}_{\mathrm{t}}) = \mathrm{V}_{\Gamma}(\bm{x})$ for all lattice translations $\bm{R}_{\mathrm{t}} \in \Gamma$. Working in the Heisenberg picture, the electronic degrees of freedom are described by a pair of electron field operators $\hat{\psi}_{s}(\bm{x},t)$ labelled by the $\mathrm{SU}(2)$ index $s \in \{\uparrow,\downarrow\}$ and satisfying
\begin{align}
    \big[\hat{\psi}_{s}(\bm{x},t), \hat{\psi}_{s'}^{\dagger}(\bm{y},t)\big]_+ &= \delta_{ss'} \delta(\bm{x}-\bm{y}),\nonumber \\
    \big[\hat{\psi}_{s}(\bm{x},t), \hat{\psi}_{s'}(\bm{y},t)\big]_+ &= \big[\hat{\psi}_{s}^{\dagger}(\bm{x},t), \hat{\psi}_{s'}^{\dagger}(\bm{y},t)\big]_+ = 0.
    \label{anticommutationrelations}
\end{align}
We assemble these into a $2$-spinor electron field operator $\hat{\psi}_0(\bm{x},t) = (\hat{\psi}_{\uparrow}(\bm{x},t), \hat{\psi}_{\downarrow}(\bm{x},t))^{\mathrm{T}}$ for which the dynamics is governed by the Heisenberg equation
\begin{align}
    i\hbar \frac{\partial \hat{\psi}_0(\bm{x},t)}{\partial t} = \big[\hat{\psi}_0(\bm{x},t), \hat{H}_0(t)\big]_-,
    \label{TDSE}
\end{align}
where the second-quantized Hamiltonian is
\begin{align}
    \hat{H}_0(t) = \int d\bm{x}\, \hat{\psi}_0^{\dagger}(\bm{x},t)\mathcal{H}_0(\bm{x})\hat{\psi}_0(\bm{x},t),
    \label{H0}
\end{align}
Here we work within the independent-particle approximation, with electron-electron interactions neglected or treated at a mean-field level as Hartree corrections to the potential energy $\mathrm{V}_{\Gamma}(\bm{x})$. The Hamiltonian density is given by
\begin{align}
    \mathcal{H}_0(\bm{x}) = \frac{1}{2m}\big(\bm{\mathfrak{p}}(\bm{x})\big)^2 + \mathrm{V}_{\Gamma}(\bm{x}) - \frac{e\hbar}{2mc} \bm{\sigma}\cdot\bm{b}_{\mathrm{static}}(\bm{x})\nonumber \\
    + \frac{\hbar}{4m^2 c^2} \bm{\sigma}\cdot\bm{\nabla}\mathrm{V}_{\Gamma}(\bm{x}) \times \bm{\mathfrak{p}}(\bm{x}),
    \label{H0density}
\end{align}
where we have defined 
\begin{align}
    \bm{\mathfrak{p}}(\bm{x}) \equiv \frac{\hbar}{i} \bm{\nabla} - \frac{e}{c} \bm{a}_{\mathrm{static}}(\bm{x}),
    \label{frakturp}
\end{align}
featuring a static vector potential which is cell-periodic, $\bm{a}_{\mathrm{static}}(\bm{x} + \bm{R}_{\mathrm{t}}) = \bm{a}_{\mathrm{static}}(\bm{x})$ for all $\bm{R}_{\mathrm{t}} \in \Gamma$. This generates a static magnetic field at the microscopic level,
\begin{align}
    \bm{b}_{\mathrm{static}}(\bm{x}) = \bm{\nabla} \times \bm{a}_{\mathrm{static}}(\bm{x}),
    \label{staticB}
\end{align}
which is also cell-periodic. This kind of effective magnetic field may arise, for example, due to local moments associated with one or more magnetic ions in each unit cell of the crystal. Notably, the first two terms of Eq. (\ref{H0density}) are to be thought of as being multiplied by the $2\times 2$ identity matrix, while the third term is the Zeeman term and the fourth term describes spin-orbit coupling; both of these feature the $3$-tuple $\bm{\sigma} = (\sigma_1, \sigma_2, \sigma_3)$ of Pauli matrices. 

Since both the electrostatic potential energy and the static vector potential are cell-periodic, so too is the Hamiltonian (\ref{H0density}). From Bloch's theorem \cite{Bloch1929}, the solutions of the spectral equation
\begin{align}
    \mathcal{H}_0(\bm{x}) \psi_{n\bm{k}}(\bm{x}) = E_{n\bm{k}} \psi_{n\bm{k}}(\bm{x})
    \label{spectralpsi}
\end{align}
are the Bloch energy eigenfunctions
\begin{align}
    \psi_{n\bm{k}}(\bm{x}) = \frac{1}{(2\pi)^{d/2}} e^{i\bm{k}\cdot\bm{x}} u_{n\bm{k}}(\bm{x}),
    \label{Blochtheorem}
\end{align}
where the cell-periodic parts $u_{n\bm{k}}(\bm{x})$ of the Bloch energy eigenfunctions, which we will refer to as \textit{Bloch functions}, are $2$-spinor wavefunctions
\begin{align}
    u_{n\bm{k}}(\bm{x}) = \left(\mqty{u_{n\bm{k}}^{\uparrow}(\bm{x}) \\[1pt] u_{n\bm{k}}^{\downarrow}(\bm{x})}\right).
    \label{Blochfunctionspin}
\end{align}
The Bloch energy eigenfunctions and these Bloch functions are labelled by a band index $n \in \mathbb{N}$, and the crystal momentum $\bm{k}$ is an element of the Brillouin zone, which is a $d$-dimensional smooth manifold $\mathrm{BZ}^d$ that is diffeomorphic to the $d$-torus \cite{Fruchart2013}.

\subsection{Chern invariants}\label{Sec:Cherninvariants}

Chern insulators are characterized by certain integer-valued invariants that we refer to collectively as \textit{Chern invariants}, which are associated with the topological properties of the crystal's electronic band structure. Two-dimensional Chern insulators are characterized by one such Chern number, while ``bulk" Chern insulators in three dimensions are characterized by a triple of Chern numbers that can be assembled into a vector-valued invariant called the \textit{Chern vector.}

\subsubsection{The Chern number in two dimensions}

We consider first a Chern insulator in $d = 2$ dimensions. Here the relevant Chern invariant is the (first) Chern number of the occupied ``valence" bands \cite{Thouless1982,Vanderbilt2018}, given by
\begin{align}
    C_{\mathcal{V}} = \frac{1}{2\pi} \sum_n f_n \int_{\mathrm{BZ}^2} d\bm{k}\, \epsilon^{ab} \partial_a \xi_{nn}^b(\bm{k}),
    \label{ChernnumberV2}
\end{align}
involving the Cartesian components ($i \in \{x,y\}$) of the non-Abelian Berry connection
\begin{align}
    \xi_{nm}^i(\bm{k}) = \frac{1}{\Omega_{\mathrm{uc}}} \int_{\Omega} d\bm{x}\, u_{n\bm{k}}^{\dagger}(\bm{x}) \frac{\partial u_{m\bm{k}}(\bm{x})}{\partial k^i},
\end{align}
where $\Omega$ is the unit cell and $\Omega_{\mathrm{uc}}$ is its volume. This expression for the Chern number features the Levi-Civita symbol with  $\epsilon^{xy} = - \epsilon^{yx} = 1$ and $\epsilon^{xx} = \epsilon^{yy} = 0$, and the filling factor $f_n = 1$ if $n$ labels an occupied valence band, and $f_n = 0$ if $n$ labels an unoccupied conduction band.

An advantage of the field theory referenced above is that one can write the Chern number in a physically transparent way in terms of the quantities in the Hamiltonian density (\ref{H0density}). This expression for the Chern number was derived in recent work \cite{Kattan2026}, and is given by
\begin{align}
    C_{\mathcal{V}} = \frac{h^2}{4\pi i} \sum_{mn} f_{nm} \int_{\mathrm{BZ}^2} \frac{d\bm{k}}{(2\pi)^2} \frac{\varepsilon^{ab} v_{nm}^a(\bm{k}) v_{mn}^b(\bm{k})}{(E_{n\bm{k}} - E_{m\bm{k}})(E_{m\bm{k}} - E_{n\bm{k}})},
    \label{microscopicChernnumberV}
\end{align}
involving the matrix elements
\begin{align}
    \bm{v}_{nm}(\bm{k}) = \frac{1}{\Omega_{\mathrm{uc}}} \int_{\Omega} d\bm{x}\, u_{n\bm{k}}^{\dagger}(\bm{x}) \bm{v}(\bm{x}) u_{m\bm{k}}(\bm{x})
    \label{velocitymatrixelements}
\end{align}
of the velocity operator 
\begin{align}
    \bm{v}(\bm{x}) = \frac{1}{m} \bm{\mathfrak{p}}(\bm{x}) + \frac{\hbar}{4m^2 c^2} \bm{\sigma}\times \bm{\nabla}\mathrm{V}_{\Gamma}(\bm{x}).
    \label{velocityoperator}
\end{align}
The expression (\ref{microscopicChernnumberV}) is well-behaved when $m = n$ as the prefactor $f_{nm} = f_{n} - f_m$ involving the filling factors $f_{n},f_{m}$ vanishes, so that the poles in the denominator do not contribute and any issues involving the ill-definition of partial derivatives at degeneracies across the Brillouin zone that can arise in Eq. (\ref{ChernnumberV2}) are not present.

\subsubsection{The Chern vector in three dimensions}

Next we consider a Chern insulator in $d = 3$ dimensions. There is a triple of first Chern numbers $C_{\mathcal{V}}^{\alpha} \in \mathbb{Z}$ indexed by $1\leq \alpha \leq 3$ characterizing the topology of the occupied valence bands over the three fundamental $2$-cycles in the three-dimensional Brillouin zone \cite{Kohmoto1992,Winkler2016,Panati2007}, 
\begin{align}
    C_{\mathcal{V}}^{\alpha} = \frac{1}{2\pi} H^{\alpha}_{\;\; a}  \sum_n f_n \int_{\mathrm{BZ}^3} d\bm{k}\, \epsilon^{abc} \partial_{b} \xi_{nn}^c(\bm{k}),
    \label{CV3Dchart}
\end{align}
involving the matrix components
\begin{align}
    H_{\;\; a}^{\alpha} = \frac{1}{2\pi} \delta^{\alpha\beta} \bm{a}_{\beta} \cdot \hat{\bm{e}}_a,
    \label{Hmatrix}
\end{align}
where $\bm{a}_{\beta}$ is a primitive lattice vector and $\hat{\bm{e}}_a$ is the $a\mathrm{th}$ Cartesian unit vector. These Chern numbers are assembled into the \textit{Chern vector}
\begin{align}
    \bm{C}_{\mathcal{V}} = \sum_{\alpha = 1}^3 C_{\mathcal{V}}^{\alpha}\, \bm{g}_{\alpha},
    \label{Chernvector}
\end{align}
which is a proper vector quantity in the sense that it does not depend on the choice of primitive lattice vectors and their associated reciprocal lattice vectors. As with the Chern number (\ref{ChernnumberV2}) in two dimensions, we  can write this Chern vector in terms of the quantities in our Hamiltonian field theory \cite{Kattan2026}, 
\begin{align}
    \bm{C}_{\mathcal{V}} = \frac{h^2}{4\pi i} \sum_{mn} f_{nm} \int_{\mathrm{BZ}^3}\frac{d\bm{k}}{(2\pi)^3}\frac{\hat{\bm{e}}_a \epsilon^{abc} v_{nm}^b(\bm{k}) v_{mn}^c(\bm{k})}{(E_{n\bm{k}} - E_{m\bm{k}})(E_{m\bm{k}} - E_{n\bm{k}})},
    \label{microscopicChernvector}
\end{align}
which again does not depend on the choice of primitive lattice vectors for the crystal's lattice. The numerator in the integrand involves the cross product $\bm{\mathfrak{v}}_{nm}(\bm{k}) \times \bm{\mathfrak{v}}_{mn}(\bm{k})$ of the vector fields associated with the velocity matrix elements. These kinds of ``global expressions" (\ref{microscopicChernnumberV}) and (\ref{microscopicChernvector}) for the Chern number and Chern vector lend themselves naturally to an analysis using \textit{ab initio} methods such as density functional theory. In principle the sums range over all bands, but in practice achieving convergence does not seem to be a serious issue \cite{Albert2026}.

\subsection{Optical response}\label{Sec:Opticalresponse}

The Chern invariants discussed in Sec.~\ref{Sec:Cherninvariants} are properties of the unperturbed band structure of a Chern insulator \cite{Mahon2023}. We now consider the scenario where there is an electromagnetic field driving the system. Then many of the quantities in Sec.~\ref{Sec:Hamiltoniantheory} need to be modified \cite{Mahon2019}. Instead of the field operator $\hat{\psi}_0(\bm{x},t)$ for the unperturbed medium, in the presence of such an electromagnetic field the electronic degrees of freedom are described by a new field operator $\hat{\psi}(\bm{x},t)$ obeying the same anticommutation relations (\ref{anticommutationrelations}), where the dynamical evolution of this field operator is governed by a Heisenberg equation with the Hamiltonian \cite{Mahon2019}
\begin{align}
    \hat{H}_{\mathrm{mc}}(t) = \int d\bm{x}\, \hat{\psi}^{\dagger}(\bm{x},t)\mathcal{H}_{\mathrm{mc}}(\bm{x},t)\hat{\psi}(\bm{x},t),
\end{align}
involving the minimal-coupling Hamiltonian density \cite{Duff1}
\begin{align}
    \mathcal{H}_{\mathrm{mc}}(\bm{x},t) = \frac{1}{2m}\left(\bm{\mathfrak{p}}(\bm{x}) - \frac{e}{c} \bm{A}(\bm{x},t)\right)^2 + e \phi(\bm{x},t) + \mathrm{V}_{\Gamma}(\bm{x})\nonumber \\
    + \frac{\hbar}{4m^2 c^2} \bm{\sigma}\cdot\bm{\nabla}\mathrm{V}_{\Gamma}(\bm{x}) \times \left(\bm{\mathfrak{p}}(\bm{x}) - \frac{e}{c} \bm{A}(\bm{x},t)\right)\nonumber \\
    - \frac{e\hbar}{2mc} \bm{\sigma}\cdot\bm{b}_{\mathrm{static}}(\bm{x}) - \frac{e\hbar}{2mc} \bm{\sigma}\cdot\bm{B}(\bm{x},t).
    \label{Hmcdensity}
\end{align}
We have neglected local-field corrections, replacing the microscopic electromagnetic field by its macroscopic counterpart, and the macroscopic Maxwell fields are related to the scalar potential $\phi(\bm{x},t)$ and vector potential $\bm{A}(\bm{x},t)$ in the usual way. 

We study the long-wavelength optical properties of Chern insulators, focusing on linear response and neglecting any dependence of response tensors on the wavevector of the Maxwell fields. Considering a Chern insulator in $d = 2$ dimensions that is modelled as a current sheet extended in the $xy$-plane, the macroscopic current density induced in this sheet by an electric field is \cite{Kattan2026}
\begin{align}
    K^i(\bm{r},\omega) = \sigma_{2d}^{i\ell}(\omega) E_{\parallel}^{\ell}(\bm{r},\omega),
    \label{sheetcurrentdensity}
\end{align}
where $\bm{r} = (x,y)$ are planar coordinates for the sheet, the Cartesian components $i,\ell \in \{x,y\}$, and $\bm{E}_{\parallel}(\bm{r},\omega)$ is the \textit{total} electric field tangential to the sheet. The two-dimensional conductivity tensor is given by
\begin{align}
    \sigma_{2d}^{i\ell}(\omega) = \sigma_{\mathrm{K}}^{i\ell}(\omega) - \frac{e^2}{2\pi\hbar} \epsilon^{i\ell} C_{\mathcal{V}}
    \label{totalconductivity2D}
\end{align}
involving the Chern number (\ref{microscopicChernnumberV}). If we instead consider a bulk Chern insulator in $d = 3$ dimensions, the macroscopic current density is \cite{Kattan2025}
\begin{align}
    J^i(\bm{x},\omega) = \sigma_{3d}^{i\ell}(\omega) E^{\ell}(\bm{x},\omega),
    \label{bulkcurrentdensity}
\end{align}
where $\bm{x} = (x,y,z)$ and the Cartesian components $i,\ell \in \{x,y,z\}$, and $\bm{E}(\bm{x},\omega)$ is again the \textit{total} electric field. The three-dimensional conductivity tensor is given by
\begin{align}
    \sigma_{3d}^{i\ell}(\omega) = \sigma_{\mathrm{K}}^{i\ell}(\omega) - \frac{e^2}{2\pi\hbar} \epsilon^{i\ell j} \hat{\bm{e}}_j \cdot\bm{C}_{\mathcal{V}},
    \label{totalconductivity3D}
\end{align}
involving the Chern vector (\ref{microscopicChernvector}). The first term in both cases is the \textit{Kubo conductivity tensor}, which describes the part of the full conductivity that depends on frequency,
\begin{widetext}
\begin{align}
    \sigma_{\mathrm{K}}^{i\ell}(\omega) = - i\omega e^2 \hbar^2 \sum_{mn} f_{nm} \int_{\mathrm{BZ}^d} \frac{d\bm{k}}{(2\pi)^d} \frac{1}{(E_{m\bm{k}} - E_{n\bm{k}})^2}\frac{v_{nm}^i(\bm{k}) v_{mn}^{\ell}(\bm{k})}{E_{m\bm{k}} - E_{n\bm{k}} - \hbar(\omega + i0^+)},
    \label{Kuboconductivity}
\end{align}
\end{widetext}
involving the velocity matrix elements (\ref{velocitymatrixelements}). Although the units of $\sigma_{\mathrm{K}}^{ij}(\omega)$ are different for $d=2$ and $d=3$, the expression (\ref{Kuboconductivity}) identifies the Kubo susceptibility in both two and three dimensions. In the static limit $(\omega \to 0)$ this Kubo conductivity tensor vanishes for both $d=2$ and $d=3$, and the remaining terms in the conductivity tensors above describe the quantum anomalous Hall effect \cite{Chang2023}.

\section{Novel optical properties}\label{Sec:Noveloptics}

When studying the long-wavelength response of three-dimensional bulk materials, it is common to introduce an effective dielectric tensor relating the macroscopic displacement field in the medium to the macroscopic electric field \cite{Jackson1999,BornWolf1999}, given by
\begin{align}
    \varepsilon^{i\ell}(\omega) = \delta^{i\ell} + \frac{4\pi i}{\omega} \sigma_{3d}^{i\ell}(\omega),
\end{align}
in terms of the three-dimensional conductivity tensor (\ref{totalconductivity3D}). The effective dielectric tensor features in the linear constitutive relations
\begin{align}
    D^i(\bm{x},\omega) &= \varepsilon^{i\ell}(\omega) E^{\ell}(\bm{x},\omega),\nonumber \\
    B^i(\bm{x},\omega) &= \delta^{i\ell} H^{\ell}(\bm{x},\omega),
    \label{constitutiverelation}
\end{align}
relating the macroscopic electric displacement and magnetic fields $(\bm{D},\bm{B})$ to the macroscopic electric and magnetic displacement fields $(\bm{E},\bm{H})$. Notably, in the long-wavelength approximation to which we restrict ourselves, there is no contribution to the second line coming from the magnetic susceptibility. In terms of our conductivity tensor (\ref{totalconductivity3D}) for a Chern insulator in three dimensions, this becomes 
\begin{align}
    \varepsilon^{i\ell}(\omega) = \delta^{i\ell} + 4\pi\chi^{i\ell}(\omega) + \frac{2ie^2}{\hbar\omega}\epsilon^{ij\ell}\hat{\bm{e}}_j\cdot\bm{C}_{\mathcal{V}},
    \label{effdielectrictensor}
\end{align}
where we have introduced a \textit{Kubo susceptibility tensor}
\begin{align}
    \chi^{i\ell}(\omega) = e^2 \hbar^2 \sum_{mn} f_{nm} \int_{\mathrm{BZ}^3} \frac{d\bm{k}}{(2\pi)^3}\frac{1}{(E_{m\bm{k}} - E_{n\bm{k}})^2}\nonumber \\
    \times \frac{v_{nm}^i(\bm{k}) v_{mn}^{\ell}(\bm{k})}{E_{m\bm{k}} - E_{n\bm{k}} - \hbar(\omega + i0^+)}.
    \label{Kubosusc}
\end{align}
This susceptibility tensor contributes to the finite-frequency response of the medium, and the third term in the dielectric tensor (\ref{effdielectrictensor}) is associated with the quantum anomalous Hall response. Very generally, we can introduce reactive and absorptive parts of this dielectric tensor, defined by \cite{lifshitz1995physical}
\begin{align}
    \mathscr{R}\big(\varepsilon^{i\ell}(\omega)\big) \equiv \frac{1}{2}\Big(\varepsilon^{i\ell}(\omega) + \big(\varepsilon^{\ell i}(\omega)\big)^*\Big),\label{reactive} \\
    \mathscr{A}\big(\varepsilon^{i\ell}(\omega)\big) \equiv \frac{1}{2i}\Big(\varepsilon^{i\ell}(\omega) - \big(\varepsilon^{\ell i}(\omega)\big)^*\Big),
    \label{absorptive}
\end{align}
such that
\begin{align}
    \varepsilon^{i\ell}(\omega) = \mathscr{R}\big(\varepsilon^{i\ell}(\omega)\big) + i \mathscr{A}\big(\varepsilon^{i\ell}(\omega)\big).
\end{align}
For time-reversal invariant systems the Chern vector vanishes, and the Kubo susceptibility tensor (\ref{Kubosusc}) can be shown to satisfy $\chi^{\ell i}(\omega) = \chi^{i\ell}(\omega)$, with the reactive and absorptive parts of $\varepsilon^{i\ell}(\omega)$ then being the real and imaginary parts of that tensor, respectively. But for a Chern insulator, having both a non-vanishing Chern number and in general $\chi^{\ell i}(\omega) \neq \chi^{i\ell}(\omega)$, we have $\varepsilon^{\ell i}(\omega) \neq \varepsilon^{i\ell}(\omega)$, and the optical properties exhibit striking differences from those of topologically trivial media. The physical meaning of these quantities in the context of material optics is discussed in Appendix \ref{Appendix:Kuboproperties}. They are of particular note at frequencies that correspond to photon energies below the bandgap, where there is no absorption, and where they may be of interest for practical applications. We refer to this frequency range as the \textit{transparency regime}, and for the rest of this paper we restrict ourselves to this regime unless otherwise noted.

\subsection{Two cases}

To illustrate these optical properties, consider Maxwell fields that have a simple variation of space, for example, 
\begin{align}
    \bm{E}(z,\omega) = \bm{E}(\omega) e^{iqz} + \mathrm{c.c.},
    \label{simplefields}
\end{align}
propagating in a medium with a fairly general response equation of the form
\begin{align}
\label{generalmodel}
    \left(\mqty{D^x \\[2pt] D^y \\[2pt] D^z}\right) = \left(\mqty{\varepsilon^{xx}(\omega) & \varepsilon^{xy}(\omega) & 0 \\[2pt] \varepsilon^{yx}(\omega) & \varepsilon^{yy}(\omega) & 0 \\[2pt]
    0 & 0 & \varepsilon^{zz}(\omega)}\right)\left(\mqty{E^x \\[2pt] E^y \\[2pt] E^z}\right),
\end{align}
where we leave the frequency dependence of the fields  
$\bm{D}(\omega)$ and $\bm{E}(\omega)$ implicit. Then we find $D^z = E^z = 0$, and from the Maxwell equations we must satisfy
\begin{align}
    q^2 \left(\mqty{E^x \\[2pt] E^y}\right) = \tilde{\omega}^2 \left(\mqty{\varepsilon^{xx}(\omega) & \varepsilon^{xy}(\omega) \\[2pt] \varepsilon^{yx}(\omega) & \varepsilon^{yy}(\omega)}\right)\left(\mqty{E^x \\[2pt] E^y}\right).
    \label{simplefieldsmatrix}
\end{align}
For a $q(\omega)$ associated with a solution of this equation, we can introduce $n(\omega) = q(\omega) / \tilde{\omega}$ as an effective refractive index. But note that it carries a sign indicating, for example if $n(\omega)$ is real, the direction of propagation (see Eq. (\ref{simplefields})).  Then Eq. (\ref{simplefieldsmatrix}) can be written as
\begin{align}
    \left(\mqty{\varepsilon^{xx}(\omega) - n(\omega)^2 & \varepsilon^{xy}(\omega) \\[2pt] \varepsilon^{yx}(\omega) & \varepsilon^{yy}(\omega) - n(\omega)^2}\right)\left(\mqty{E^x \\[2pt] E^y}\right) = 0.
    \label{nsquaredmatrix}
\end{align}
The condition that the determinant vanishes gives
\begin{align}
    n(\omega)^2 =&\; \frac{1}{2} \big(\varepsilon^{xx}(\omega) + \varepsilon^{yy}(\omega)\big)\nonumber \\
    &\pm \frac{1}{2} \sqrt{(\varepsilon^{xx}(\omega) - \varepsilon^{yy}(\omega))^2 + 4\varepsilon^{xy}(\omega)\varepsilon^{yx}(\omega)}.
    \label{nsquared}
\end{align}
Focusing on the transparency regime, the absorptive part (\ref{absorptive}) of $\varepsilon^{i\ell}(\omega)$ vanishes, and $\varepsilon^{i\ell}(\omega)$ is purely reactive. It follows that the components $\varepsilon^{xx}(\omega)$, $\varepsilon^{yy}(\omega)$, and $\varepsilon^{xy}(\omega)\varepsilon^{yx}(\omega)$ are all purely real, and that the latter is non-negative, so both possible values of $n(\omega)^2$ are purely real (see Appendix \ref{Appendix:Kuboproperties}). Here and below we define the square root of positive real numbers $R_+$ such that $\sqrt{R_+}>0$.

We assume that both $\varepsilon^{xx}(\omega)$ and $\varepsilon^{yy}(\omega)$ are nonnegative, as expected for insulators at frequencies below the bandgap. There are then two possibilities: One value of $n(\omega)^2$ could be positive and one negative (case I), or both could be positive (case II). 
\\

\noindent \textbf{Case I:} Suppose that while 
\begin{align}
    n_{\mathrm{prop}}(\omega)^2 \equiv&\; \frac{1}{2} \big(\varepsilon^{xx}(\omega) + \varepsilon^{yy}(\omega)\big)\nonumber \\
    &+ \frac{1}{2} \sqrt{(\varepsilon^{xx}(\omega) - \varepsilon^{yy}(\omega))^2 + 4\varepsilon^{xy}(\omega)\varepsilon^{yx}(\omega)}
\end{align}
is purely real, with $n_{\mathrm{prop}}(\omega)$ taken to be positive, 
\begin{align}
    (i\kappa(\omega))^2 \equiv&\; \frac{1}{2} \big(\varepsilon^{xx}(\omega) + \varepsilon^{yy}(\omega)\big)\nonumber \\
    &-  \frac{1}{2} \sqrt{(\varepsilon^{xx}(\omega) - \varepsilon^{yy}(\omega))^2 + 4\varepsilon^{xy}(\omega)\varepsilon^{yx}(\omega)}
\end{align}
is negative, with $\kappa(\omega)$ taken to be positive. Then we can have the solutions $n(\omega) = \pm n_{\mathrm{prop}}(\omega)$ corresponding to waves propagating along the $z$ direction or the $-z$ direction with refractive index $n_{\mathrm{prop}}(\omega)$, or we can have the solutions $n(\omega) = \pm i\kappa(\omega)$ corresponding to waves that are evanescent in the $z$ direction or in the $-z$ direction. In the latter case, if we had a medium with a surface at the $z = 0$ plane, then no energy on it would be transmitted into the medium and all the light would be reflected. This is an analog of total internal reflection, but at normal incidence. We discuss this in more detail for a Chern insulator subject to light at sufficiently long wavelengths in Sec. \ref{Sec:Chernmirror}. 
\\

\noindent \textbf{Case II:} Suppose instead that both possible values (\ref{nsquared}) of $n(\omega)^2$ are positive. Then the possible values of its roots are $n(\omega) = \pm n^+(\omega)$, where
\begin{align}
    n^+(\omega) \equiv&\; \bigg[\frac{1}{2} \big(\varepsilon^{xx}(\omega) + \varepsilon^{yy}(\omega)\big)\nonumber \\
    &+ \frac{1}{2} \sqrt{(\varepsilon^{xx}(\omega) - \varepsilon^{yy}(\omega))^2 + 4\varepsilon^{xy}(\omega)\varepsilon^{yx}(\omega)}\bigg]^{1/2},
    \label{nplusgeneral}
\end{align}
or $n(\omega) = \pm n^{-}(\omega)$, where
\begin{align}
    n^-(\omega) \equiv&\; \bigg[\frac{1}{2} \big(\varepsilon^{xx}(\omega) + \varepsilon^{yy}(\omega)\big)\nonumber \\
    &- \frac{1}{2} \sqrt{(\varepsilon^{xx}(\omega) - \varepsilon^{yy}(\omega))^2 + 4\varepsilon^{xy}(\omega)\varepsilon^{yx}(\omega)}\bigg]^{1/2}.
    \label{nminusgeneral}
\end{align}
That is, we can have waves propagating in the $z$ direction and in the $-z$ direction, and in either case they can be propagating with refractive indices $n^+(\omega)$ or $n^-(\omega)$. This can lead to optical activity, but of a type rather different than the ``natural" optical activity exhibited by solutions of chiral molecules or crystals such as $\alpha$--quartz, where there is a chiral ordering of the $\mathrm{Si}\mathrm{O}_2$ molecular components. To highlight the difference, we first recall the natural optical activity that those media illustrate.

\subsection{Natural optical activity}

Natural optical activity is a special case of a spatially dispersive response of the medium, where the response depends on the wavevector $\bm{q}$ of the Maxwell fields \cite{Landau1984,Lekner1996}. The results here are well-known, but we review some of the basic features of the optics here to contrast with that of a Chern insulator. An effective dielectric tensor $\varepsilon_{\mathrm{eff}}^{i\ell}(\bm{q},\omega)$ relating the components of $\bm{D}(\bm{q},\omega)$ to $\bm{E}(\bm{q},\omega)$ can be introduced, and it can be written as
\begin{align}
    \varepsilon_{\mathrm{eff}}^{i\ell}(\bm{q},\omega) = \varepsilon^{i\ell}(\omega) + \epsilon^{i\ell j} g^{jk}(\omega) q^k,
\end{align}
where $g^{\ell k}(\omega)$ is a second-rank axial tensor (the ``gyration tensor"), and $\varepsilon^{i\ell}(\omega)$ is the $\bm{q} = \bm{0}$ limit of this effective dielectric tensor. For $\alpha$--quartz, for example, where $\varepsilon^{i\ell}(\omega)$ is diagonal and $\varepsilon^{xx}(\omega) = \varepsilon^{yy}(\omega) \equiv \varepsilon^{\perp}(\omega)$, the point group is $32$, the gyration tensor takes the form $\bm{g}(\omega) = \mathrm{diag}(g^{11}(\omega), g^{22}(\omega), g^{33}(\omega))$ if the $z$ axis is chosen to lie along the $c$ axis of the crystal, and for light propagating with a wavevector $\bm{q} = q\hat{\bm{e}}_z$ only $g^{33}(\omega)$ contributes \cite{Nye1985, Landau1984}; writing it just as $g(\omega)$, we have
\begin{align}
    \varepsilon_{\mathrm{eff}}^{i\ell}(\bm{q},\omega) = \left(\mqty{\varepsilon^{\perp}(\omega) & ig(\omega)q & 0 \\[2pt] -ig(\omega)q & \varepsilon^{\perp}(\omega) & 0 \\[2pt]
    0 & 0 & \varepsilon^{zz}(\omega)}\right).
\end{align}
Again, for fields of the form (\ref{simplefields}), from Maxwell's equations propagating fields with only $x$ and $y$ components will then satisfy Eq. (\ref{simplefieldsmatrix}), or
\begin{align}
    \left(\mqty{\tilde{\omega}^2\varepsilon^{\perp}(\omega) - q^2 & i\tilde{\omega}^2 g(\omega)q \\[2pt] -i\tilde{\omega}^2 g(\omega)q & \tilde{\omega}^2\varepsilon^{\perp}(\omega) - q^2}\right)\left(\mqty{E^x \\[2pt] E^y}\right) = 0.
    \label{qmatrixeqn}
\end{align}
The dispersion relations of propagating fields then follow from setting the determinant of the matrix to zero,
\begin{align}
    \big(\tilde{\omega}^2 \varepsilon^{\perp}(\omega) - q^2\big) = \pm \tilde{\omega}^2 g(\omega) q. 
    \label{conditionq}
\end{align}
Defining the effective refractive index $n(\omega) = q / \tilde{\omega}$, from the choice of positive sign we find
\begin{align}
    n(\omega) = \pm \sqrt{\varepsilon^{\perp}(\omega) + \frac{1}{4}\tilde{\omega}^2 g(\omega)^2} - \frac{1}{2}\tilde{\omega} g(\omega),
    \label{neffectiveqpositive}
\end{align}
while from the choice of negative sign we find
\begin{align}
    n(\omega) = \pm \sqrt{\varepsilon^{\perp}(\omega) + \frac{1}{4}\tilde{\omega}^2 g(\omega)^2} + \frac{1}{2}\tilde{\omega} g(\omega).
    \label{neffectiveqnegative}
\end{align}
So as expected there are two indices, 
\begin{align}
    n^{\pm}(\omega) = \sqrt{\varepsilon^{\perp}(\omega) + \frac{1}{4}\tilde{\omega}^2 g(\omega)^2} \pm \frac{1}{2}\tilde{\omega} g(\omega),
\end{align}
and there are waves propagating forward and backward with both of these indices. That is, we have four solutions $n(\omega) = \pm n^+(\omega)$ and $n(\omega) = \pm n^-(\omega)$; we assume that $g(\omega)>0$ so indeed $n^+(\omega)>n^-(\omega)$.

\begin{figure}[h]
    \centering
    \includegraphics[width=\linewidth]{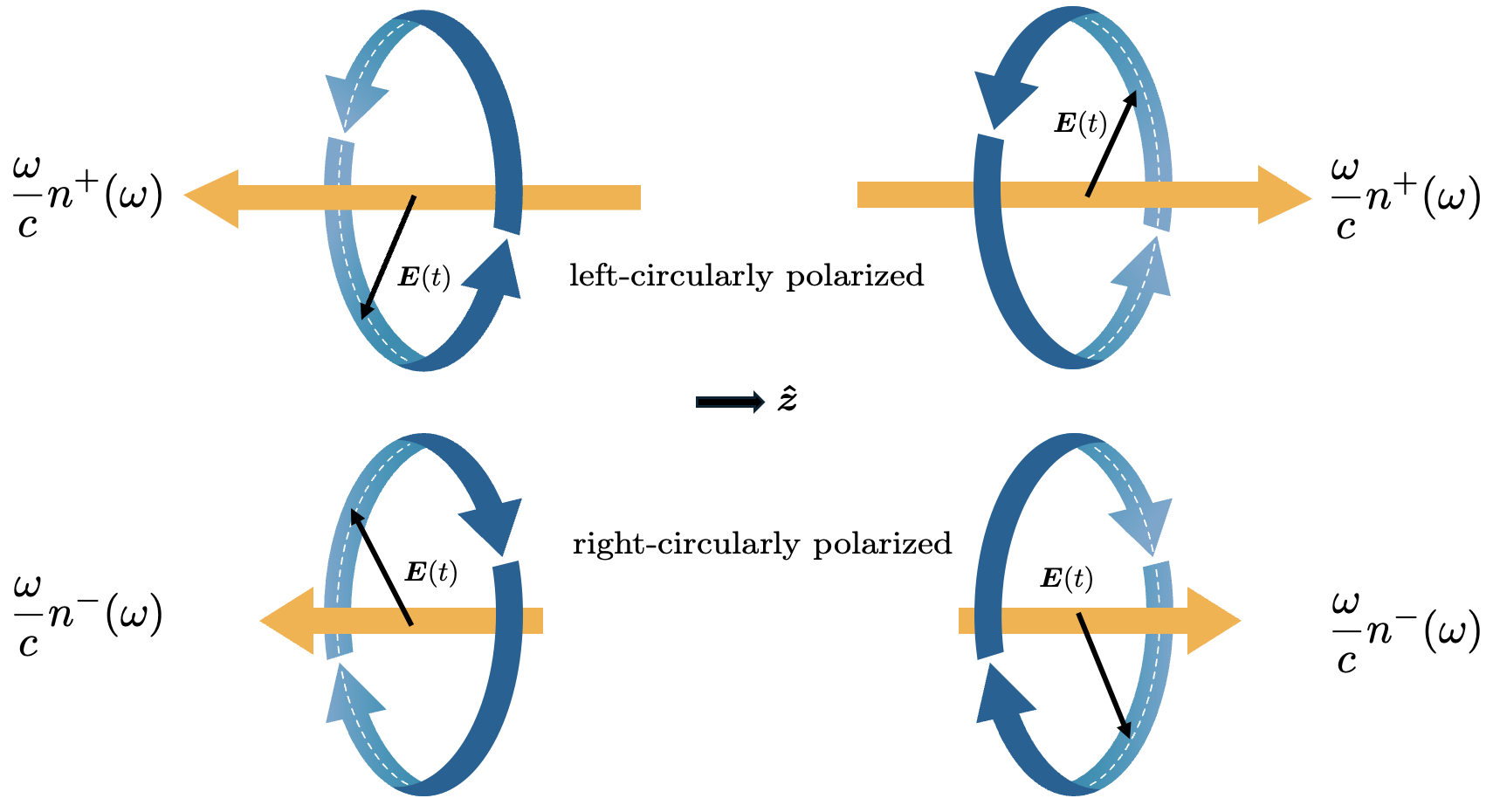}
    \caption{The refractive indices in natural optical activity do not depend on the direction of propagation, but only on the polarization.}
    \label{fig:naturaloptact}
\end{figure}

Consider first light propagating with index $n^+(\omega)$ in the $z$ direction; this situation results from the top sign in Eq. (\ref{neffectiveqpositive}) and thus the positive sign in Eq. (\ref{conditionq}), and from Eq. (\ref{qmatrixeqn}) we see that corresponds to propagation with $E^y / E^x = - i$; since $q > 0$, this corresponds to left-handed circularly polarized light. For light propagating with index $n^+(\omega)$ in the $-z$ direction, the situation results from the bottom sign of Eq. (\ref{neffectiveqnegative}) and thus the negative sign in Eq. (\ref{conditionq}), and from Eq. (\ref{qmatrixeqn}) we see that corresponds to propagation with $E^y / E^x = i$; since $q < 0$, this also corresponds to left-handed circularly polarized light. So light travelling with index $n^+(\omega)$ is left-circularly polarized whether it is propagating along the $z$ direction or the $-z$ direction, and light travelling with index $n^-(\omega)$ is right-circularly polarized whether it is propagating along the $z$ direction or the $-z$ direction. See Figure~\ref{fig:naturaloptact}.

\subsection{``Unnatural" optical activity}

The result for a bulk Chern insulator is strikingly different, and can be traced back to the fact that the optical activity is not due to spatial dispersion effects, but instead to the unusual form of the response tensor $\varepsilon^{i\ell}(\omega)$. Since it arises in the long-wavelength approximation, it can be expected to be a larger effect than natural optical activity, which relies on spatial dispersion. 

To compare directly with the situation for $\alpha$--quartz, we simplify our earlier expressions by putting
\begin{align}
    \varepsilon^{xx}(\omega) = \varepsilon^{yy}(\omega) \equiv \varepsilon^{\perp}(\omega).
\end{align}
Then the solutions (\ref{nsquared}) for $n(\omega)^2$ reduce to 
\begin{align}
    n(\omega)^2 = \varepsilon^{\perp}(\omega) \pm \sqrt{\varepsilon^{xy}(\omega)\varepsilon^{yx}(\omega)},
\end{align}
and the equation for $E^x$ and $E^y$ simplifies to
\begin{align}
    \left(\mqty{\varepsilon^{\perp}(\omega) - n(\omega)^2 & \varepsilon^{xy}(\omega) \\[2pt] \varepsilon^{yx}(\omega) & \varepsilon^{\perp}(\omega) - n(\omega)^2}\right)\left(\mqty{E^x \\[2pt] E^y}\right) = 0.
\end{align}
For $n(\omega) = \pm n^+(\omega)$, where here
\begin{align}
    n^+(\omega) = \sqrt{\varepsilon^{\perp}(\omega) + \sqrt{\varepsilon^{xy}(\omega)\varepsilon^{yx}(\omega)}},
    \label{Nplus}
\end{align}
we have
\begin{align}
    \left(\mqty{- \sqrt{\varepsilon^{xy}(\omega)\varepsilon^{yx}(\omega)} & \varepsilon^{xy}(\omega) \\[2pt] \varepsilon^{yx}(\omega) & - \sqrt{\varepsilon^{xy}(\omega)\varepsilon^{yx}(\omega)}}\right)\left(\mqty{E^x \\[2pt] E^y}\right) = 0,
\end{align}
and so
\begin{align}
    \frac{E^y}{E^x} = \frac{\varepsilon^{yx}(\omega)}{\sqrt{\varepsilon^{xy}(\omega)\varepsilon^{yx}(\omega)}},
    \label{Eratioplus}
\end{align}
while for $n(\omega) = \pm n^-(\omega)$, where here
\begin{align}
    n^-(\omega) = \sqrt{\varepsilon^{\perp}(\omega) - \sqrt{\varepsilon^{xy}(\omega)\varepsilon^{yx}(\omega)}},
    \label{Nminus}
\end{align}
we have
\begin{align}
    \left(\mqty{\sqrt{\varepsilon^{xy}(\omega)\varepsilon^{yx}(\omega)} & \varepsilon^{xy}(\omega) \\[2pt] \varepsilon^{yx}(\omega) & \sqrt{\varepsilon^{xy}(\omega)\varepsilon^{yx}(\omega)}}\right)\left(\mqty{E^x \\[2pt] E^y}\right) = 0,
\end{align}
and so
\begin{align}
    \frac{E^y}{E^x} = - \frac{\varepsilon^{yx}(\omega)}{\sqrt{\varepsilon^{xy}(\omega)\varepsilon^{yx}(\omega)}}.
    \label{Eratiominus}
\end{align}
Note that the equations (\ref{Eratioplus}) and (\ref{Eratiominus}) that identify the polarization states depend only on $\varepsilon^{xy}(\omega)$ and $\varepsilon^{yx}(\omega)$. In the transparency regime, on which we focus, we have $(n^+(\omega))^2$ and $(n^-(\omega))^2$ both positive, but whether we have $n^+(\omega)$ and $n^-(\omega)$ both real (Case II) or we have $n^+(\omega)$ real and $n^-(\omega)$ imaginary (Case I) depends also on the size of $\varepsilon^{\perp}(\omega)$ as well.

Now for a three-dimensional Chern insulator we have
\begin{align}
    \varepsilon^{xy}(\omega) &= 4\pi \chi^{xy}(\omega) - \frac{2ie^2}{\hbar\omega} C_{\mathcal{V}}^z,\nonumber \\
    \varepsilon^{yx}(\omega) &= 4\pi \chi^{yx}(\omega) + \frac{2ie^2}{\hbar\omega} C_{\mathcal{V}}^z,
    \label{epsilons}
\end{align}
where $C_{\mathcal{V}}^z \equiv \hat{\bm{e}}_z\cdot\bm{C}_{\mathcal{V}}$ is the Cartesian projection of the Chern vector along the direction of propagation. In general, the off-diagonal components of $\varepsilon^{i\ell}(\omega)$ will have both symmetric and antisymmetric parts, and so Eqs. (\ref{Eratioplus}) and (\ref{Eratiominus}) will generally correspond to elliptically polarized light. But we can directly compare with $\alpha$--quartz for a Chern insulator subject to light at a low enough frequency that the off-diagonal contribution to $\varepsilon^{i\ell}(\omega)$ from $\chi^{i\ell}(\omega)$ is negligible; this would be expected for a Chern insulator at sufficiently low frequencies, where the contribution from the Chern vector projection will dominate. Then the off-diagonal components of the dielectric tensor become purely ``topological," namely 
\begin{align}
    \varepsilon^{xy}(\omega) &\to - \frac{2ie^2}{\hbar\omega} C_{\mathcal{V}}^z,\nonumber \\
    \varepsilon^{yx}(\omega) &\to \frac{2ie^2}{\hbar\omega} C_{\mathcal{V}}^z,
\end{align}
and the indices become
\begin{align}
    n^{\pm}(\omega) \to \sqrt{\varepsilon^{\perp}(\omega) \pm \frac{2e^2}{\hbar\omega} \abs{C_{\mathcal{V}}^z}}.
    \label{ns}
\end{align}
Again, for comparison with $\alpha$-quartz we assume that $\varepsilon^{\perp}(\omega)$ is sufficiently large that $n^{\pm}(\omega)$ are real (Case II).

\begin{figure}[h]
    \centering
    \includegraphics[width=\linewidth]{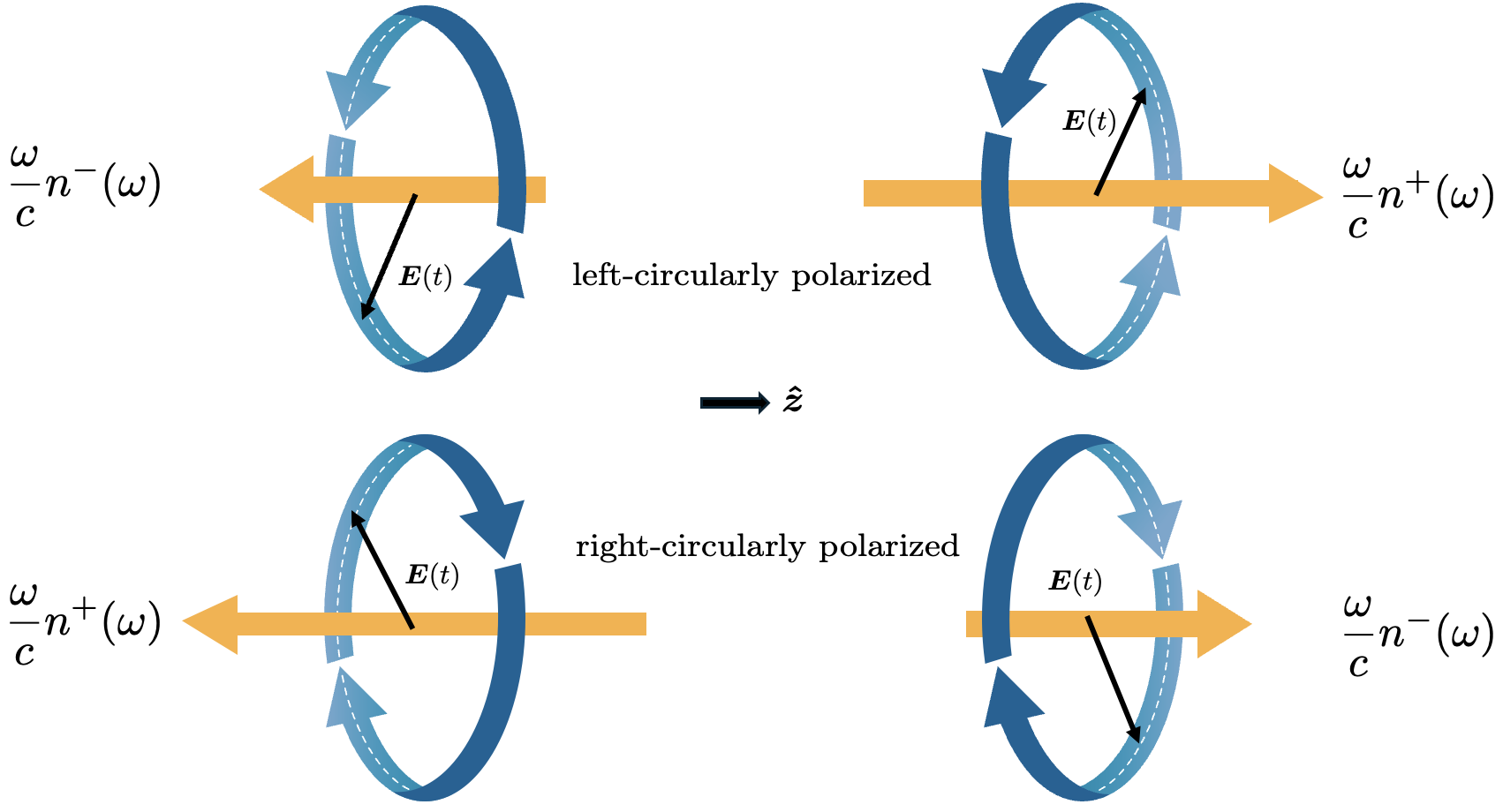}
    \caption{The refractive indices in ``unnatural" optical activity depend on both the direction of propagation and the polarization. The diagrams are drawn for $\nu>0$, and in drawing them we have assumed that both $n^+(\omega)$ and $n^-(\omega)$ are real.}
    \label{fig:unnaturaloptact}
\end{figure}

Now for $n(\omega) = \pm n^+(\omega)$ we see from Eq. (\ref{Eratioplus}) that
\begin{align}
    \frac{E^y}{E^x} = i \frac{C_{\mathcal{V}}^z}{\abs{C_{\mathcal{V}}^z}},
\end{align}
where if $C_{\mathcal{V}}^z < 0$ this corresponds to left-circularly polarized light if it is propagating in the $z$ direction with index $n(\omega) = n^+(\omega)$, and to right-circularly polarized light if it is propagating in the $-z$ direction with index $n(\omega) = - n^+(\omega)$. And for $n(\omega) = \pm n^-(\omega)$ we see from Eq. (\ref{Eratiominus}) that 
\begin{align}
    \frac{E^y}{E^x} = - i \frac{C_{\mathcal{V}}^z}{\abs{C_{\mathcal{V}}^z}},
\end{align}
which if $C_{\mathcal{V}}^z < 0$ corresponds to right-circularly polarized light if it is propagating along the $z$ direction with index $n(\omega) = n^-(\omega)$, and to left-circularly polarized light if it is propagating along the $-z$ direction with index $n(\omega) = - n^-(\omega)$. While in ``naturally" optically active media light of a given handedness has the same index of refraction whether it is propagating in the $z$ direction or the $-z$ direction, here light of a given handedness has a \textit{different} index of refraction whether it is propagating in the $z$ direction or the $-z$ direction. And if $C_{\mathcal{V}}^z >0$ the circularity of the light propagating in a given direction with a given index is reversed compared with the above, but the qualitative feature that light of a given handedness has a different index depending on its direction of propagation still holds. See Figure~\ref{fig:unnaturaloptact}. 

To be more general, we can allow for $\varepsilon^{\perp}(\omega)$ to be small enough that $n^{-}(\omega)$ becomes purely imaginary (Case I). Furthermore, at higher frequencies the off-diagonal components $\chi^{xy}(\omega)$ and $\chi^{yx}(\omega)$ 
of the Kubo susceptibility tensor (\ref{Kubosusc}) can be significant. Very generally, if we just restrict ourselves to the transparency regime, only the reactive part (\ref{reactive}) of the dielectric tensor can exist. For crystals that have a $C_{nz}$ in-plane point-group symmetry with $n > 2$, one can show that $\chi^{xy}(\omega) = -\chi^{yx}(\omega)$, and then the off-diagonal components of the dielectric tensor must be purely imaginary; as well, $\varepsilon^{xx}(\omega) = \varepsilon^{yy}(\omega)$, and this quantity, which we again write as $\varepsilon^{\perp}(\omega)$, will be purely real (see Appendix \ref{Appendix:Kuboproperties}). So again the circularity of the light propagating with a given index is reversed if the direction is reversed, although which direction is associated with which index for a given polarization will depend on the size of the components contributing to $\varepsilon^{xy}(\omega) = -\varepsilon^{yx}(\omega)$ in  Eq. (\ref{epsilons}). This can be encoded in the sign function
\begin{align}
    \nu(\omega) = \mathrm{sgn}\left(4\pi\Im\chi^{xy}(\omega) - \frac{2e^2}{\hbar\omega}C_{\mathcal{V}}^z\right),
    \label{nu}
\end{align}
which has the property $\nu(\omega) \to - \mathrm{sgn}(C_{\mathcal{V}}^z)$ when the off-diagonal components of the Kubo susceptibility are negligible. The circularity of upward and downward propagating fields corresponding to the two indices and values of $\nu(\omega) = \pm 1$ are shown in Table \ref{Table:Refractiveindices}, and the indices $n^{\pm}(\omega)$, which in the general response model of Eq. (\ref{generalmodel}) follow from Eq. (\ref{nsquared}), are here given by Eqs. (\ref{Nplus}) and (\ref{Nminus}). Here either $n^+(\omega)$ will be real and $n^-(\omega)$ will be imaginary (Case I), or $n^+(\omega)$ and $n^-(\omega)$ will both be real (Case II).

\begin{table}[h]
\label{tab:nu_polarizations}
\begin{ruledtabular}
\begin{tabular}{c c c c}
Refractive index & Propagation & $\nu=-1$ & $\nu=+1$ \\[0.3em]
\hline
$n^+(\omega)$ & $+z$ & right-handed & left-handed \\[0.8em]
$n^+(\omega)$ & $-z$ & left-handed & right-handed \\[0.8em]
$n^-(\omega)$ & $+z$ & left-handed & right-handed \\[0.8em]
$n^-(\omega)$ & $-z$ & right-handed & left-handed
\end{tabular}
\caption{The circular polarization components corresponding to the refractive indices $n^{\pm}(\omega)$ for light propagating along the $\pm z$ directions.} 
\label{Table:Refractiveindices}
\end{ruledtabular}
\end{table}

\section{Bulk optics}\label{Sec:Bulkoptics}

With these results in hand, we consider a bulk Chern insulator of finite thickness that is extended in the $xy$-plane, where we imagine that this thickness is sufficiently large so that a bulk description in three dimensions is valid. We restrict ourselves to a scenario described at the end of the last section: We are in the transparency regime, and the symmetry is high enough that for the normal incidence problems we consider the indices $n^{\pm}(\omega)$ are given by Eqs. (\ref{Nplus}) and (\ref{Nminus}), and the polarization of waves propagating (or evanescent) upward or downward are all circular and given by Table I. 

We proceed as in studying the optics of thin films of simpler materials.  Fresnel coefficients are introduced for the different interfaces, and they are then combined to describe the transmission and reflection due to the entire structure.  With these in hand we can evaluate the Faraday and Kerr angles that describe the unnatural optical activity in a bulk Chern insulator, and show that they are topologically quantized at sufficiently long wavelengths.

Only normal incidence is considered here; we plan to treat more general excitation scenarios in an upcoming contribution.

\subsection{Transmission and reflection in three dimensions}\label{Sec:Transmissionreflection3d}

\begin{figure}[h]
\centering
\includegraphics[width=7cm]{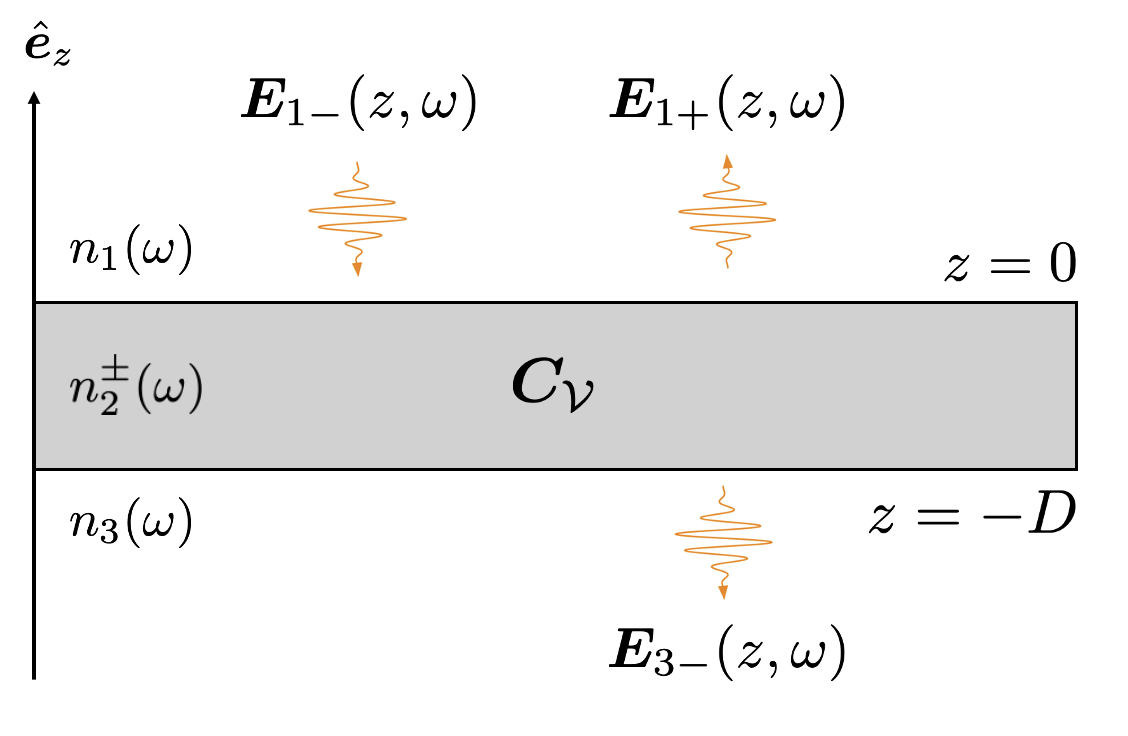}
\caption{Here we consider a three-dimensional Chern insulator of finite thickness $D > 0$ with Chern vector (\ref{microscopicChernvector}), where the regions $z > 0$ and $z < 0$ are isotropic media with indices of refraction $n_1(\omega)$ and $n_3(\omega)$, respectively. The transmitted field is $\bm{E}_{3-}(z,\omega)$ and the reflected field is $\bm{E}_{1+}(z,\omega)$ with circular polarization components for which the propagation thereof is described by the bulk refractive indices $n_2^{\pm}(\omega)$ for the Chern insulator.}
\label{Slabgeometry}
\end{figure}

The setup we consider is shown in Fig.~\ref{Slabgeometry}: We take there to be light propagating along the $z$ direction in a system consisting of a three-dimensional Chern insulator of thickness $D$ in the region $- D < z < 0$ (region $2$) that is extended in the $xy$-plane, bounded in region $1$ above ($z > 0$) by an isotropic medium such as air, and in the region $3$ below $(z < - D)$ by some other isotropic medium such as an insulating or metallic substrate. The isotropic media in the regions $i = 1,3$ each have a single dielectric function $\varepsilon_i(\omega)$ with an associated index of refraction $n_i(\omega) = \sqrt{\varepsilon_i(\omega)}$, and the fields propagating inside the medium move at a phase velocity characterized by the indices (\ref{nplusgeneral}) and (\ref{nminusgeneral}).

The Maxwell fields across each of the boundary interfaces are related by a set of Fresnel coefficients for the left and right circularly polarized components of these fields. For an electric field $\bm{E}_{1-}(z,\omega)$ propagating downwards along the $z$ direction at normal incidence on the interface $z = 0$, the Fresnel coefficients for this interface are defined by the relations
\begin{align}
    E_{2-}^{\alpha}(z=0,\omega) &= t_{12}^{\beta\alpha}(\omega) E_{1-}^{\beta}(z=0,\omega),\nonumber \\
    E_{1+}^{\alpha}(z=0,\omega) &= r_{12}^{\beta\alpha}(\omega) E_{1-}^{\beta}(z=0,\omega),
    \label{Fresnel0}
\end{align}
where the indices $\alpha,\beta = L, R$ with the repeated index $\beta$ indicating a sum; ``$L$" indicates left-handed circularly polarized light, and ``$R$" indicates right-handed circularly polarized light. The first line involves the transmitted field $\bm{E}_{2-}(z,\omega)$ and the second line involves the reflected field $\bm{E}_{1+}(z,\omega)$, both of which are evaluated at the interface $z = 0$. These Fresnel coefficients can be written in terms of the refractive indices for the media above and below this interface using the Maxwell \textit{saltus} conditions \cite{Jackson1999} that enforce continuity of the tangential electric and magnetic fields across the interface. The nonvanishing Fresnel coefficients are found to be
\begin{align}
    t_{12}^{LL}(\omega) &= \frac{2 n_{1}(\omega)}{n_{1}(\omega) + n_{2}^{\nu}(\omega)},\;\;\; t_{12}^{RR}(\omega) = \frac{2 n_{1}(\omega)}{n_{1}(\omega) + n_{2}^{-\nu}(\omega)}, \nonumber \\
    r_{12}^{LR}(\omega) &= \frac{n_{1}(\omega) - n_{2}^{\nu}(\omega)}{n_{1}(\omega) + n_{2}^{\nu}(\omega)},\;\;\; r_{12}^{RL}(\omega) = \frac{n_{1}(\omega) - n_{2}^{-\nu}(\omega)}{n_{1}(\omega) + n_{2}^{-\nu}(\omega)}, 
    \label{Fresnel12}
\end{align}
where the superscripts on the indices of the Chern insulator indicate the sign function (\ref{nu}). Similarly, the Fresnel coefficients for the interface $z = - D$ are defined by
\begin{align}
    E_{3-}^{\alpha}(z=-D,\omega) &= t_{23}^{\beta\alpha}(\omega) E_{2-}^{\beta}(z=-D,\omega),\nonumber \\
    E_{2+}^{\alpha}(z=-D,\omega) &= r_{23}^{\beta\alpha}(\omega) E_{2-}^{\beta}(z=-D,\omega),
    \label{FresnelD}
\end{align}
involving the transmitted field $\bm{E}_{3-}(z,\omega)$ in the first line and the reflected field $\bm{E}_{2+}(z,\omega)$ in the second line evaluated at this interface. The nonvanishing Fresnel coefficients are found to be
\begin{align}
    t_{23}^{LL}(\omega) &= \frac{2 n_{2}^{\nu}(\omega)}{n_{2}^{\nu}(\omega) + n_{3}(\omega)},\;\;\; t_{23}^{RR}(\omega) = \frac{2 n_{2}^{-\nu}(\omega)}{n_{2}^{-\nu}(\omega) + n_{3}(\omega)}, \nonumber \\
    r_{23}^{LR}(\omega) &= \frac{n_{2}^{\nu}(\omega) - n_{3}(\omega)}{n_{2}^{\nu}(\omega) + n_{3}(\omega)},\;\;\; r_{23}^{RL}(\omega) = \frac{n_{2}^{-\nu}(\omega) - n_{3}(\omega)}{n_{2}^{-\nu}(\omega) + n_{3}(\omega)}.
    \label{Fresnel23}
\end{align}
If the medium in region $2$ is topologically trivial, then the refractive indices $n_2^{+}(\omega) = n_2^{-}(\omega)$ and the Fresnel coefficients at both interfaces reduce to the usual expressions for an interface between isotropic media \cite{Jackson1999}.

While the Fresnel coefficients relate the Maxwell fields across the boundary interfaces of the sample, the fields across the entire sample in region $2$ of Fig.~\ref{Slabgeometry} are related by a collection of \textit{bulk} transmission and reflection coefficients, which are defined by the relations
\begin{align}
    E_{3-}^{\alpha}(z=-D,\omega) &= \mathcal{T}_{13}^{\beta\alpha}(\omega) E_{1-}^{\beta}(z=0,\omega),\nonumber \\
    E_{1+}^{\alpha}(z=0,\omega) &= \mathcal{R}_{13}^{\beta\alpha}(\omega) E_{1-}^{\beta}(z=0,\omega),
    \label{bulktensors}
\end{align}
involving the transmitted field $\bm{E}_{3-}(z,\omega)$ at the interface $z = - D$ in the first line and the reflected field $\bm{E}_{1+}(z,\omega)$ at the interface $z = 0$ in the second line. Following a lengthy calculation in terms of transfer matrices that is relegated to Appendix \ref{Appendix:Bulktransmissionreflection}, we find the bulk transmission coefficients
\begin{align}
    \mathcal{T}_{13}^{LL}(\omega) &= \frac{ t_{12}^{LL}(\omega) t_{23}^{LL}(\omega) e^{i\tilde{\omega}n_2^{\nu} D}}{1 - r_{23}^{LR}(\omega) r_{21}^{RL}(\omega) e^{2i\tilde{\omega} n_2^{\nu} D}},\nonumber \\
    \mathcal{T}_{13}^{RR}(\omega) &= \frac{t_{12}^{RR}(\omega) t_{23}^{RR}(\omega) e^{i\tilde{\omega}n_2^{-\nu} D}}{1 - r_{23}^{RL}(\omega) r_{21}^{LR}(\omega) e^{2i\tilde{\omega} n_2^{-\nu} D}},\nonumber \\
    \mathcal{T}_{13}^{LR}(\omega) &= \mathcal{T}_{13}^{RL}(\omega) = 0,
    \label{bulktransmission}
\end{align}
and the bulk reflection coefficients
\begin{align}
    \mathcal{R}_{13}^{LR}(\omega) &= r_{12}^{LR}(\omega) + \frac{t_{12}^{LL}(\omega) r_{23}^{LR}(\omega) t_{21}^{RR}(\omega) e^{2i\tilde{\omega}n_2^{\nu} D}}{1 - r_{23}^{LR}(\omega) r_{21}^{RL}(\omega) e^{2i\tilde{\omega} n_2^{\nu} D}},\nonumber \\
    \mathcal{R}_{13}^{RL}(\omega) &= r_{12}^{RL}(\omega) + \frac{t_{12}^{RR}(\omega) r_{23}^{RL}(\omega) t_{21}^{LL}(\omega) e^{2i\tilde{\omega}n_2^{-\nu} D}}{1 - r_{23}^{RL}(\omega) r_{21}^{LR}(\omega) e^{2i\tilde{\omega} n_2^{-\nu} D}},\nonumber \\
    \mathcal{R}_{13}^{LL}(\omega) &= \mathcal{R}_{13}^{RR}(\omega) = 0.
    \label{bulkreflection}
\end{align}
Note that these expressions are valid only when considering light at normal incidence. They are the generalization to anisotropic media of the Airy formulas for transmission and reflection of light across a simple dielectric \cite{Airyformulaeref}.

\subsection{Optical signatures of topology}

We have seen how the anisotropy of the conductivity tensor (\ref{totalconductivity3D}) leads to a pair of distinct refractive indices that describe the phase velocities of left and right circularly polarized light propagating in the medium. This means that a Chern insulator is optically active, and its unnatural optical activity is characterized by complex-valued Faraday and Kerr phases representing the differential phase acquired by the transmitted and reflected fields, respectively, as they propagate across the sample. For the scenario 
of Fig.~\ref{Slabgeometry}, these phases are taken to be
\begin{align}
    \delta_F^{\mathrm{bulk}}(\omega) &= - \frac{i}{2} \mathrm{Log}\left(\frac{\mathcal{T}_{13}^{LL}(\omega)}{\mathcal{T}_{13}^{RR}(\omega)}\right),\nonumber \\
    \delta_K^{\mathrm{bulk}}(\omega) &= - \frac{i}{2} \mathrm{Log}\left(\frac{\mathcal{R}_{13}^{LR}(\omega)}{\mathcal{R}_{13}^{RL}(\omega)}\right),
    \label{FaradayKerrphases}
\end{align}
involving the principal (complex) logarithm of ratios of the bulk transmission and reflection coefficients, respectively \cite{Chakraborty2023FaradayKerr}. The real parts of these phases, called the \textit{Faraday angle} and \textit{Kerr angle}, respectively, are associated with circular birefringence, where the plane of polarization of the light rotates as it propagates across the medium, and are given by
\begin{align}
    \theta_F^{\mathrm{bulk}}(\omega) &= \frac{1}{2} \mathrm{Arg}\left(\frac{\mathcal{T}_{13}^{LL}(\omega)}{\mathcal{T}_{13}^{RR}(\omega)}\right),\nonumber \\
    \theta_K^{\mathrm{bulk}}(\omega) &= \frac{1}{2} \mathrm{Arg}\left(\frac{\mathcal{R}_{13}^{LR}(\omega)}{\mathcal{R}_{13}^{RL}(\omega)}\right),
    \label{FaradayKerrangle}
\end{align}
where $\mathrm{Arg}(z)$ denotes the principal argument that lies in the interval $(-\pi,\pi]$. Meanwhile, the imaginary parts of these phases are associated with circular dichroism, where the left and right circularly polarized components of light suffer different amounts of absorption. One usually works with
\begin{align}
    \tanh\eta_F^{\mathrm{bulk}}(\omega) &= \frac{|\mathcal{T}_{13}^{RR}(\omega)| - |\mathcal{T}_{13}^{LL}(\omega)|}{|\mathcal{T}_{13}^{LL}(\omega)| + |\mathcal{T}_{13}^{RR}(\omega)|},\nonumber \\
    \tanh\eta_K^{\mathrm{bulk}}(\omega) &= \frac{|\mathcal{R}_{13}^{RL}(\omega)| - |\mathcal{R}_{13}^{LR}(\omega)|}{|\mathcal{R}_{13}^{LR}(\omega)| + |\mathcal{R}_{13}^{RL}(\omega)|},
    \label{FaradayKerrellipticity}
\end{align}
which are called the \textit{Faraday ellipticity} and \textit{Kerr ellipticity}, respectively, because they describe the conversion of circularly polarized light into elliptically polarized light under transmission and reflection, respectively \cite{Chakraborty2023FaradayKerr}.

Since there have not yet been any three-dimensional materials found to support a true ``bulk" Chern insulator phase, we will not provide a numerical calculation of these quantities here; this will be done for their two-dimensional analogues in Sec.~\ref{Sec:Sheetoptics}. But it is interesting to consider the behaviour of these quantities in the very-long-wavelength limit. Suppose that there is eventually found to be some material with minimal direct bandgap $E_{\mathrm{gap}}$ that supports a bulk Chern insulator phase. For light with photon energies $\hbar\omega \ll E_{\mathrm{gap}}$, we can Taylor expand the integrand of the Kubo susceptibility (\ref{Kubosusc}) in powers of $\hbar\omega/E_{\mathrm{gap}}$, and thereby calculate the leading-order contributions to the Faraday and Kerr effects. Following a straightforward calculation, the Faraday angle for the transmitted field takes the form
\begin{align}
    \theta_F^{\mathrm{bulk}}(\omega) \simeq - \arctan(\frac{2\alpha D C_{\mathcal{V}}^z}{n_1(\omega) + n_3(\omega)}),
    \label{BulkFaradayquantization}
\end{align}
while the Kerr angle for the reflected field takes the form
\begin{align}
    \theta_K^{\mathrm{bulk}}(\omega) \simeq &- \arctan(\frac{2\alpha D C_{\mathcal{V}}^z}{n_1(\omega) + n_3(\omega)})\nonumber \\
    &- \arctan(\frac{2\alpha D C_{\mathcal{V}}^z}{n_1(\omega) - n_3(\omega)}),
    \label{BulkKerrquantization}
\end{align}
where $\alpha = e^2/\hbar c$ is the fine-structure constant. These expressions involve the principal real inverse tangent, with values taken in the interval $(-\pi/2,\pi/2)$, and ``$\simeq$" indicates an equality up to terms of order $O(\hbar\omega/E_{\mathrm{gap}})$. The Faraday and Kerr ellipticities vanish to this same order. The Faraday and Kerr angles characterize the orientation of the plane of linear polarization, and, since this plane is unchanged by a rotation through $\pi$, the Faraday and Kerr angles are only defined modulo $\pi$. Thus, when $n_1(\omega)=n_3(\omega)=1$ the second term in the Kerr angle is understood by continuity modulo $\pi$. And in this limit of a bulk Chern insulator in vacuum where $n_1(\omega)=n_3(\omega)=1$, the Faraday and Kerr angles become ``topologically quantized" in the sense that they depend only on the Chern vector $\bm{C}_{\mathcal{V}}$, and the thickness $D$ of the sample. We will return to these Faraday and Kerr effects in a two-dimensional Chern insulator in Sec.~\ref{Sec:Sheetoptics}.

\section{The Chern mirror}\label{Sec:Chernmirror}

Consider next a scenario where there is a planar interface forming the boundary between two bulk Chern insulators. The purely DC response at this kind of interface is the topic of the ``bulk-boundary correspondence" \cite{Halperin1982,Chang2023}, which states that the transport along the boundary is unidirectional and is determined by the difference of Chern invariants for the bulk Chern insulators above and below the boundary. It is interesting to see what an analysis of the optical properties of such a boundary reveals, especially for optical wavelengths sufficiently large that only the topological terms in Eq. (\ref{epsilons}) contribute.

We model this system as consisting of a bulk Chern insulator with Chern vector $\bm{C}_{\mathcal{V}_1}$ in region $1$ with $z > 0$, and another bulk Chern insulator with Chern vector $\bm{C}_{\mathcal{V}_2}$ in region $2$ with $z < 0$. Of course, all Chern insulators are finite-sized, but we may treat the two media as being semi-infinite when their thicknesses are large enough so that the light reflected from their remote surfaces is either attenuated or incoherent by the time it returns to the interface; then the optical response is governed only by the interface $z = 0$. We will be interested in the transmission and reflection of light across this interface, which is described by the Fresnel coefficients
\begin{align}
    t_{12}^{LL}(\omega) &= \frac{2 n_{1}^{\nu}(\omega)}{n_{1}^{\nu}(\omega) + n_{2}^{\nu}(\omega)},\;\;\; t_{12}^{RR}(\omega) = \frac{2 n_{1}^{-\nu}(\omega)}{n_{1}^{-\nu}(\omega) + n_{2}^{-\nu}(\omega)}, \\
    r_{12}^{LR}(\omega) &= \frac{n_{1}^{\nu}(\omega) - n_{2}^{\nu}(\omega)}{n_{1}^{\nu}(\omega) + n_{2}^{\nu}(\omega)},\;\;\; r_{12}^{RL}(\omega) = \frac{n_{1}^{-\nu}(\omega) - n_{2}^{-\nu}(\omega)}{n_{1}^{-\nu}(\omega) + n_{2}^{-\nu}(\omega)}.
    \label{generalFresnelcoeffs}
\end{align}
These expressions describe the transmission and reflection of a general, frequency-dependent incident field at normal incidence on this interface. However, at sufficiently long wavelengths corresponding to frequencies $\omega$ with $\hbar\omega \ll E_{\mathrm{gap}}$, where $E_{\mathrm{gap}}$ is the smaller of the two minimal direct bandgaps of the two bulk Chern insulators, the response is dominated by the third term in the dielectric tensor (\ref{effdielectrictensor}). In this very-long-wavelength limit, the Fresnel coefficients become
\begin{align}
    t_{12}^{LL}(\omega) &= t_{12}^{RR}(\omega) \simeq \frac{2\sqrt{C_{1}^z}}{\sqrt{C_1^z} + \sqrt{C_2^z}},\nonumber \\
    r_{12}^{LR}(\omega) &= r_{12}^{RL}(\omega) \simeq \frac{\sqrt{C_1^z} - \sqrt{C_2^z}}{\sqrt{C_1^z} + \sqrt{C_2^z}},
    \label{staticlimit2}
\end{align}
involving the Chern projections $C_i^z \equiv \hat{\bm{e}}_z \cdot \bm{C}_{\mathcal{V}_i}$ ($i = 1,2$), and the square roots are taken in the principal branch. As before, the notation ``$\simeq$" indicates an equality up to terms of order $O(\hbar\omega/E_{\mathrm{gap}})$. 

We consider an incident field $\bm{E}_{1-}(z,\omega)$ in region $1$ propagating along the $-z$ direction, and there will be a reflected field $\bm{E}_{1+}(z,\omega)$ in this region propagating along the $z$ direction, along with a transmitted field $\bm{E}_{2-}(z,\omega)$ in region $2$ propagating along the $-z$ direction. In this scenario, it is interesting to consider the \textit{power transmission and reflection coefficients} 
\begin{align}
    T_{12}(\omega) \equiv \frac{\big|\hat{\bm{e}}_z \cdot \big\langle\bm{S}_{2-}(z = 0,\omega)\big\rangle\big|}{\big|\hat{\bm{e}}_z \cdot \big\langle\bm{S}_{1-}(z = 0,\omega)\big\rangle\big|},\nonumber \\
    R_{12}(\omega) \equiv \frac{\big|\hat{\bm{e}}_z \cdot \big\langle\bm{S}_{1+}(z = 0,\omega)\big\rangle\big|}{\big|\hat{\bm{e}}_z \cdot \big\langle\bm{S}_{1-}(z = 0,\omega)\big\rangle\big|},
    \label{PowerTRdefinition}
\end{align}
involving the time-averaged Poynting vectors
\begin{align}
    \big\langle \bm{S}_{i\pm}(z,\omega)\big\rangle = \frac{c}{2\pi} \Re\Big(\bm{E}_{i\pm}(z,\omega) \times \bm{B}_{i\pm}^*(z,\omega)\Big).
\end{align}
These quantities depend on the moduli of the generally complex-valued Fresnel coefficients (\ref{staticlimit2}). There are two distinct cases, depending on the relative signs of the Chern projections above and below the interface $z = 0$. 

We first consider the scenario where both Chern projections have the same sign; that is, we assume both Chern projections are positive or both are negative. When both Chern projections are positive, $C_{1}^z > 0$ and $C_{2}^z > 0$, we find that right-circularly polarized light propagating in the $-z$ direction in both regions is oscillatory, while left-circularly polarized light propagating in the $-z$ direction in both regions is evanescent. The circularity is reversed for light propagating along the $z$ direction. Similarly, when both Chern projections are negative, $C_{1}^z < 0$ and $C_{2}^z < 0$, we find that left-circularly polarized light propagating along the $-z$ direction in both regions is oscillatory, while right-circularly polarized light propagating along the $-z$ direction in both regions is evanescent. Again, the circularity is reversed for light propagating along the $z$ direction. Then the power transmission and reflection coefficients (\ref{PowerTRdefinition}) are 
\begin{align}
    T_{12}(\omega) &\simeq \frac{4\sqrt{\abs{C_{1}^z C_{2}^z}}}{(\sqrt{\abs{C_1^z}} + \sqrt{\abs{C_2^z}})^2},\nonumber \\
    R_{12}(\omega) &\simeq \left(\frac{\sqrt{\abs{C_1^z}} - \sqrt{\abs{C_2^z}}}{\sqrt{\abs{C_1^z}} + \sqrt{\abs{C_2^z}}}\right)^2,
    \label{TRsamesign}
\end{align}
up to terms of order $O(\hbar\omega/E_{\mathrm{gap}})$. This situation involves a partial transmission and a partial reflection of the incident power at the interface $z = 0$, a result of the mismatch between the topologically quantized Hall impedances across this interface.

Next we consider the scenario where the Chern projections have opposite signs, that is, when one of the Chern projections is positive and the other Chern projection is negative. For example, when $C_{1}^z > 0$ and $C_{2}^z < 0$ we find that right-circularly polarized light propagating in the $-z$ direction in region $1$ is oscillatory and left-circularly polarized light is evanescent, while the opposite holds in region $2$. The circularity is reversed for light propagating along the $z$ direction. In the other case where $C_{1}^z < 0$ and $C_{2}^z > 0$, we find that left-circularly polarized light propagating in the $-z$ direction in region $1$ is oscillatory and right-circularly polarized light is evanescent, while the opposite holds in region $2$. Again, the circularity is reversed for light propagating along the $z$ direction. These two cases together seem to imply that there is no transmitted power across the interface $z = 0$ when the signs of the Chern projections differ, and, moreover, this statement seems to be independent of the magnitude of these Chern projections. And, indeed, the power transmission and reflection coefficients (\ref{PowerTRdefinition}) are
\begin{align}
    T_{12}(\omega) &\simeq 0,\nonumber \\
    R_{12}(\omega) &\simeq 1,
    \label{TRoppositesign}
\end{align}
up to terms of order $O(\hbar\omega/E_{\mathrm{gap}})$. This is a remarkable result because it does not depend at all on the magnitude of the Chern projections, and only relies on them having opposite signs! It is a kind of ``topological" total internal reflection, but it occurs at normal incidence. Because this effect does not occur for other ``conventional" topological insulators in the tenfold way -- in those materials the Chern invariants vanish due to time-reversal symmetry -- it is perhaps best referred to as a kind of \textit{Chern mirror effect}. Indeed, analogous to a standard metallic mirror, the transmitted field is purely evanescent. But this holds in an \textit{insulator}. %and occurs no matter how large the difference in the absolute value of the Chern projections across the interface. 
This is also very different than the ordinary total internal reflection that occurs in the optics of topologically trivial media \cite{Jackson1999}, which requires non-normal incidence and a difference in the magnitude of the relevant impedances. This is yet another manifestation of the ``unnatural" optical activity of Sec.~\ref{Sec:Noveloptics}.

\section{Sheet optics}\label{Sec:Sheetoptics}

We have discussed several manifestations of ``unnatural" optical activity in three-dimensional bulk Chern insulators. However, all of the current material platforms for Chern insulators are effectively two-dimensional thin films \cite{Yu2010,Chang2013,Chang2015,Liu2020,Li2024}, and to demonstrate how this strange form of optical activity arises in such materials, we consider the optical properties of a Chern insulator in two dimensions. This models the scenario where the thickness of the film is sufficiently small that the incident field is no longer cancelled in the medium by extinction, and instead the film behaves like a current sheet that sources a radiated field above and below the film, along with the incident field that exists in both regions.

\subsection{Transmission and reflection in two dimensions}

The material geometry we consider is a thin film of a Chern insulator that is extended in the $xy$-plane and is placed just above the substrate medium with index $n_3(\omega)$ that is taken to extend up to $z=0^-$; there is no medium with indices $n_2^{\pm}(\omega)$, and the medium in which the sheet is embedded is vacuum, $n_1(\omega)=1$. Then the macroscopic current density for this sheet that enters as a source in the Maxwell equations is 
\begin{align}
    \bm{J}(\bm{x},\omega) = \delta(z) \bm{K}(\bm{r},\omega),
    \label{volumetricJ}
\end{align}
where $\bm{r} = (x,y)$ are planar coordinates for the sheet, and $\bm{K}(\bm{r},\omega)$ is the sheet current density (\ref{sheetcurrentdensity}) involving the two-dimensional conductivity tensor (\ref{totalconductivity2D}). The region $1$ above the sheet ($z > 0$) will be air ($n_1(\omega) = 1$), and the substrate region $3$ below the sheet ($z < 0$) will be a simple dielectric with index $n_3(\omega)$. 

The Maxwell fields are related across the sheet by a pair of sheet transmission and reflection tensors; these are the two-dimensional analogues of the bulk transmission and reflection coefficients of Sec.~\ref{Sec:Bulkoptics}. These are the transmission and reflection coefficients for a sheet at $z = 0$ embedded in region $1$, modified by the presence of a substrate in region $3$ below. Working in the circular polarization basis, the components of these sheet transmission and reflection coefficients are defined by the relations
\begin{align}
    E_{3-}^{\alpha}(0^-,\omega) &= \mathfrak{t}_{13}^{\beta\alpha}(\omega) E_{1-}^{\beta}(0^+,\omega),\nonumber \\
    E_{1+}^{\alpha}(0^+,\omega) &= \mathfrak{r}_{13}^{\beta\alpha}(\omega) E_{1-}^{\beta}(0^+,\omega),
\end{align}
where $\alpha,\beta = L,R$ and the repeated index $\beta$ is summed. Here $\bm{E}_{3-}(0^-,\omega)$ is the downward propagating transmitted field just below the sheet, while $\bm{E}_{1+}(0^+,\omega)$ is the upward propagating reflected field just above the sheet. The calculation of these coefficients in terms of transfer matrices is summarized in Appendix \ref{Appendix:Sheettransfermatrix}, where we find for the transmission coefficients
\begin{align}
    \mathfrak{t}_{13}^{LL}(\omega) &= \frac{t_{-}^{LL}(\omega) t_{13}(\omega)}{1 - r_{13}(\omega) r_{+}^{RL}(\omega)},\nonumber \\
    \mathfrak{t}_{13}^{RR}(\omega) &= \frac{t_{-}^{RR}(\omega)t_{13}(\omega)}{1 - r_{13}(\omega) r_{+}^{LR}(\omega)},\nonumber \\
    \mathfrak{t}_{13}^{LR}(\omega) &= \mathfrak{t}_{13}^{RL}(\omega) = 0.
    \label{Sheettransmissioncoeffs}
\end{align}
and for the reflection coefficients
\begin{align}
    \mathfrak{r}_{13}^{LR}(\omega) &= r_{-}^{LR}(\omega) + \frac{t_{-}^{LL}(\omega) r_{13}(\omega) t_{+}^{RR}(\omega)}{1 - r_{13}(\omega)r_{+}^{RL}(\omega)},\nonumber \\
    \mathfrak{r}_{13}^{RL}(\omega) &= r_{-}^{RL}(\omega) + \frac{t_{-}^{RR}(\omega) r_{13}(\omega) t_{+}^{LL}(\omega)}{1 - r_{13}(\omega) r_{+}^{LR}(\omega)},\nonumber \\
    \mathfrak{r}_{13}^{LL}(\omega) &= \mathfrak{r}_{13}^{RR}(\omega) = 0.
    \label{Sheetreflectioncoeffs}
\end{align}
Here the first superscript labels the incident polarization and the second labels the outgoing polarization, while the subscript ``$+$" or ``$-$" labels the propagation direction of the incident field. Thus, incidence from region 1 above the sheet corresponds to the subscript ``$-$," and \textit{vice versa}. These coefficients feature two sets of quantities, the first of which describe the transmission and reflection of light across the sheet embedded only in dielectric in region $1$, 
\begin{align}
    t_{\pm}^{LL}(\omega) &= \frac{1}{1 + \frac{2\pi}{c} \Sigma^{\pm}(\omega)},\nonumber \\
    t_{\pm}^{RR}(\omega) &= \frac{1}{1 + \frac{2\pi}{c} \Sigma^{\mp}(\omega)},\nonumber \\
    r_{\pm}^{LR}(\omega) &= - \frac{\frac{2\pi}{c}\Sigma^{\pm}(\omega)}{1+ \frac{2\pi}{c} \Sigma^{\pm}(\omega)},\nonumber \\
    r_{\pm}^{RL}(\omega) &= - \frac{\frac{2\pi}{c}\Sigma^{\mp}(\omega)}{1 + \frac{2\pi}{c} \Sigma^{\mp}(\omega)},
\end{align}
involving the helicity-channel conductivities
\begin{align}
    \Sigma^{\pm}(\omega) \equiv \sigma_{2d}^{xx}(\omega) \pm i\sigma_{2d}^{xy}(\omega).
\end{align}
The sheet transmission and reflection coefficients also feature the ordinary Fresnel coefficients for the $1$-to-$3$ interface, given by
\begin{align}
    t_{13}(\omega) = \frac{2}{1 + n_3(\omega)}, \quad r_{13}(\omega) = \frac{1 - n_3(\omega)}{1 + n_3(\omega)},
\end{align}
involving the index $n_3(\omega)$ %are the indices 
for the dielectric substrate.

\subsection{``Unnatural" optical activity in $\mathrm{Mn}\mathrm{Bi}_2\mathrm{Te}_4$}

We illustrate the huge enhancement in circular birefringence and circular dichroism that arises in a material Chern insulator because of the long-wavelength character of ``unnatural" optical activity. By analogy with Sec.~\ref{Sec:Bulkoptics}, the Faraday and Kerr angles describing circular birefringence in a two-dimensional current sheet are
\begin{align}
    \theta_F^{\mathrm{sheet}}(\omega) &= \frac{1}{2} \mathrm{Arg}\left(\frac{\mathfrak{t}_{13}^{LL}(\omega)}{\mathfrak{t}_{13}^{RR}(\omega)}\right),\nonumber \\
    \theta_K^{\mathrm{sheet}}(\omega) &= \frac{1}{2} \mathrm{Arg}\left(\frac{\mathfrak{r}_{13}^{LR}(\omega)}{\mathfrak{r}_{13}^{RL}(\omega)}\right),
    \label{FaradayKerranglesheet}
\end{align}
and the Faraday and Kerr ellipticities describing circular dichroism in the current sheet are 
\begin{align}
    \tanh\eta_F^{\mathrm{sheet}}(\omega) &= \frac{|\mathfrak{t}_{13}^{RR}(\omega)| - |\mathfrak{t}_{13}^{LL}(\omega)|}{|\mathfrak{t}_{13}^{LL}(\omega)| + |\mathfrak{t}_{13}^{RR}(\omega)|},\nonumber \\
    \tanh\eta_K^{\mathrm{sheet}}(\omega) &= \frac{|\mathfrak{r}_{13}^{RL}(\omega)| - |\mathfrak{r}_{13}^{LR}(\omega)|}{|\mathfrak{r}_{13}^{LR}(\omega)| + |\mathfrak{r}_{13}^{RL}(\omega)|},
    \label{FaradayKerrellipticity}
\end{align}
involving the sheet transmission (\ref{Sheettransmissioncoeffs}) and reflection (\ref{Sheetreflectioncoeffs}) coefficients for the multilayer structure. 

We take as our sheet Chern insulator a thin film of $\mathrm{Mn}\mathrm{Bi}_2\mathrm{Te}_4$ with $N = 5$ septuple layers, which below the critical temperature has a Chern number $C_{\mathcal{V}} = - 1$ \cite{Li2024}. We consider light with energies that are below the bandgap of $57\,\mathrm{meV}$ for this system so that absorption can be neglected. This system has the necessary $C_{3z}$ in-plane point-group symmetry that is required for light propagating therein to be circularly polarized. We consider the experimentally realistic scenario with air in region $1$ and a substrate in region $3$ with a single, possibly complex-valued refractive index $n_3(\omega)$. We calculate the Faraday and Kerr angles for light with various energies in the THz and far-infrared regimes using numerical data for the conductivity tensor in other work \cite{Albert2026}. The values for these angles with silicon ($\mathrm{Si}$) and silicon dioxide ($\mathrm{Si}\mathrm{O}_2$) substrates are summarized in Table~\ref{Table:Angles}.

\begin{table}[h]
    \centering
    \begin{tabular}{|p{1.75cm}||p{1.25cm}|p{1.25cm}||p{1.25cm}|p{1.25cm}|}
        \hline
        \multirow{2}{*}{Energy}
        & \multicolumn{2}{c||}{Faraday angle $\theta_F^{\mathrm{sheet}}(\omega)$}
        & \multicolumn{2}{c|}{Kerr angle $\theta_K^{\mathrm{sheet}}(\omega)$} \\
        \cline{2-5}
        & $\mathrm{Si}$ & $\mathrm{SiO}_2$
        & $\mathrm{Si}$ & $\mathrm{SiO}_2$ \\
        \hline
        $1$ meV  & $0.190^{\circ}$ & $0.283^{\circ}$ & $-0.159^{\circ}$ & $-0.593^{\circ}$ \\
        $10$ meV & $0.193^{\circ}$ & $0.286^{\circ}$ & $-0.161^{\circ}$ & $-0.595^{\circ}$ \\
        $20$ meV & $0.200^{\circ}$ & $0.293^{\circ}$ & $-0.167^{\circ}$ & $-0.592^{\circ}$ \\
        $30$ meV & $0.213^{\circ}$ & $0.306^{\circ}$ & $-0.178^{\circ}$ & $-0.579^{\circ}$ \\
        $40$ meV & $0.239^{\circ}$ & $0.328^{\circ}$ & $-0.200^{\circ}$ & $-0.544^{\circ}$ \\
        $50$ meV & $0.302^{\circ}$ & $0.371^{\circ}$ & $-0.251^{\circ}$ & $-0.454^{\circ}$ \\
        \hline
    \end{tabular}
    \caption{Faraday and Kerr angles for the transmitted and reflected fields after propagating across an $N = 5$ septuple layer thin film of $\mathrm{Mn}\mathrm{Bi}_2\mathrm{Te}_4$ on various substrates.}
    \label{Table:Angles}
\end{table}

It should be emphasized that the Faraday and Kerr angles in this table are extremely large compared with, for example, a sample of $\alpha$-quartz of the same thickness, which is optically active but topologically trivial; there the Faraday and Kerr angles are found to be on the order of $10^{-5}$ degrees \cite{Lekner1996}, and so these angles are roughly five orders of magnitude larger. This massive enhancement of the Faraday and Kerr effects has been experimentally observed in other topological insulators \cite{Maciejko2010,Tse2011,Wu2016}, including the axion $(\mathbb{Z}_2)$ topological insulator $\mathrm{Bi}_2\mathrm{Se}_3$ where Faraday and Kerr angles were observed at around the same order of magnitude in the $\mathrm{THz}$ regime. But while both $\mathbb{Z}_2$ topological insulators and Chern insulators have a similar enhancement of the Faraday and Kerr effects in this regime, the behaviour of these quantities is very different in Chern insulators for light at microwave and RF wavelengths, where the topological quantization of the Chern invariant plays a central role. Indeed, for light at these wavelengths the Faraday and Kerr angles become ``topologically quantized." When the light has energies satisfying $\hbar\omega \ll E_{\mathrm{gap}}$, where $E_{\mathrm{gap}}$ is around 30--160 meV for all known material Chern insulators, the Faraday angle becomes
\begin{align}
    \theta_F^{\mathrm{sheet}}(\omega) = - \arctan(\frac{2\alpha C_{\mathcal{V}}}{1 + n_3(\omega)}) + O\left(\frac{\hbar\omega}{E_{\mathrm{gap}}}\right).
    \label{thetaF}
\end{align}
and the Kerr angle becomes
\begin{align}
    \theta_K^{\mathrm{sheet}}(\omega) = &- \arctan(\frac{2\alpha C_{\mathcal{V}}}{1 + n_3(\omega)})\nonumber \\
    &- \arctan(\frac{2\alpha C_{\mathcal{V}}}{1 - n_3(\omega)}) + O\left(\frac{\hbar\omega}{E_{\mathrm{gap}}}\right).
    \label{thetaK}
\end{align}
where $\alpha = e^2/\hbar c$ is the fine-structure constant; the Faraday and Kerr ellipticities vanish. The inverse tangents and limiting values are understood according to the same convention as in Eqs. (\ref{BulkFaradayquantization}) and (\ref{BulkKerrquantization}). For an isolated sheet Chern insulator where $n_3(\omega)=1$, in the limit $\hbar\omega/E_{\mathrm{gap}} \ll 1$ we have
\begin{align}
     \theta_F^{\mathrm{sheet}}(\omega) = - \arctan(\alpha C_{\mathcal{V}})
\end{align}
and
\begin{align}
    \theta_K^{\mathrm{sheet}}(\omega) = \mathrm{sgn}(C_{\mathcal{V}})\left(\frac{\pi}{2} - \arctan(\alpha \abs{C_{\mathcal{V}}})\right).
\end{align}
Note that Eqs. (\ref{thetaF}) and (\ref{thetaK}) are analogous to Eqs. (\ref{BulkFaradayquantization}) and (\ref{BulkKerrquantization}) for a bulk Chern insulator, and vanish for topologically trivial systems.

\section{Discussion}\label{Sec:Discussion}

We have demonstrated that the topological properties of a Chern insulator's band structure can be probed in a direct way using optics. The essential link between the material properties and their long-wavelength response to light is provided by a conductivity tensor that was derived in recent work \cite{Kattan2025}, in which the relevant Chern invariant in two or three dimensions enters as a part of the optical response that survives in the static limit. The anisotropy of this conductivity tensor, both in the Hall term and in the Kubo term that describes the part of the optical response that vanishes in the static limit, results in a kind of ``unnatural" optical activity that is fundamentally different than the natural optical activity that arises in crystals such as $\alpha$-quartz, and in solutions of chiral molecules.

Natural optical activity is one type of spatially dispersive response that depends on the wavevector of the Maxwell fields \cite{Landau1984}, and vanishes in the long-wavelength limit. It is characterized by a pair of refractive indices for left-handed and right-handed circularly polarized light; for a given refractive index the handedness of the propagation is unchanged if the direction of propagation of the light is reversed. But the unnatural optical activity that occurs in a Chern insulator is different. It survives in the long-wavelength limit, and the identification of each of the refractive indices with the handedness of the light actually depends on the direction of propagation of light in the Chern insulator. Fundamentally, this is an optical manifestation of broken time-reversal symmetry, and also of the quantum anomalous Hall response, which is topological in its origin. 

Once this directional asymmetry is recognized, the novel optical properties of Chern insulators become more clear. We first considered light at normal incidence on a bulk Chern insulator that is extended in two spatial dimensions and of finite thickness, where this thickness was assumed to be sufficiently large that a discussion in terms of refractive indices is valid. In Sec.~\ref{Sec:Bulkoptics}, we studied the transmission and reflection of light across this system, which is characterized by a collection of bulk transmission and reflection coefficients that are a generalization of the usual Airy formulas for isotropic dielectrics \cite{Jackson1999}. In terms of these bulk coefficients we derived the Faraday and Kerr angles, and the corresponding ellipticities, which in the absence of absorption characterize circular birefringence and circular dichroism, respectively. At sufficiently long wavelengths, the Faraday and Kerr angles become topologically quantized, depending on the projection of the Chern vector along the direction of propagation of the light, while the Faraday and Kerr ellipticities were found to vanish. In this sense, the Faraday and Kerr angles provide us with a direct optical probe of the topological properties of a Chern insulator: The rotation is fixed by the Chern invariant that characterizes its topology, dressed only by the refractive indices of the media on either side of the sample.  

The same unnatural optical activity leads to a strange ``Chern mirror effect" at the interface between two bulk Chern insulators; this was the topic of Sec.~\ref{Sec:Chernmirror}. At this interface, the sign of the Chern vector projected along the direction of propagation determines which circular polarization channel is propagating, and which channel is evanescent, at least for light at sufficiently long wavelengths. If the two Chern projections have the same sign, then a propagating channel exists on both sides of the interface, and the problem resembles that of an ordinary mismatch of impedances, except that the impedances are set by the topological Hall response. But when the two Chern projections have opposite signs, the propagating channel on one side is paired with an evanescent channel on the other. No propagating optical channel is then available to carry power across the interface, and the incident power is completely reflected. This is analogous to total internal reflection, in the sense that the transmitted field is evanescent, but it is qualitatively different in origin: It occurs at normal incidence, in a band insulator, and depends only on the relative signs of the Chern invariants.

For all of the currently known materials that support a Chern insulator phase \cite{Deng2020,Liu2020}, such as thin films of magnetically doped $\mathrm{Bi}_2\mathrm{Te}_3$ and odd-number stacks of $\mathrm{Mn}\mathrm{Bi}_2\mathrm{Te}_4$ septuple layers, a more realistic geometry is that of a two-dimensional current sheet, instead of a three-dimensional bulk medium. However, the transmission and reflection calculations of Sec.~\ref{Sec:Sheetoptics}, where we modelled these thin films as a current sheet with a Chern number instead of a Chern vector, demonstrate that the same physics associated with this unnatural optical activity survives in the thin-film limit. The Chern number enters the sheet conductivity as a Hall conductance, and the multilayer optical problem can again be written in terms of circular polarization channels. While there is no ``propagation" within a current sheet, the sheet transmission and reflection coefficients can be written in a form similar to that of the coefficients for the bulk, and from these sheet coefficients the Faraday and Kerr angles, and Faraday and Kerr ellipticities, could be defined. We calculated these quantities for a thin film of $\mathrm{Mn}\mathrm{Bi}_2\mathrm{Te}_4$ with five septuple layers, where the quantum anomalous Hall effect has been observed experimentally, and found there to be a massive enhancement of these quantities as compared with the naturally optically active $\alpha$-quartz, for example. This enhancement reflects the fact that the unnatural optical activity of a Chern insulator is a long-wavelength effect, and is not suppressed by a factor of the wavelength of light as in the spatially dispersive optical activity. And, as in the bulk case, we showed that at sufficiently long wavelengths the topological term again dominates, resulting in a polarization rotation that is controlled by Chern number. 

We have restricted ourselves here to light at normal incidence, and at frequencies below the band gap where there is no absorption, because then the circular polarization basis diagonalizes the effective dielectric tensor whenever there is sufficient in-plane crystal symmetry. At non-normal incidence, or at frequencies above the band gap where there is absorption, the eigenmodes will be elliptically polarized in general, and we expect the interplay between this ellipticity and the anisotropic character of the response of a Chern insulator to be even more interesting; this is a topic for future work. Focused beams and other structured fields provide another promising direction, since the finite distribution of wavevectors in a beam will probe the topological response beyond that of a single mode. More broadly, multilayer structures containing Chern-insulator sheets offer a route to engineering topological optical elements with a polarization response controlled by Chern number, magnetic order, and dielectric environment. The central lesson is clear: The topological properties of Chern insulators are not confined to DC transport, and they can profoundly influence the behaviour of light. This makes the optical properties of Chern insulators a promising route to probing band topology without relying on the explicit edge-state picture of the ``bulk-boundary correspondence" \cite{Halperin1982}, and suggests that engineered multilayer geometries could be used to amplify, suppress, or otherwise control the polarization of light in these topological materials.

%%%%%%%%%%%%%%%%%%%%%%%%%%%%%%%%%%%%%%%%%%

\begin{acknowledgments}
    This work was supported by the Natural Sciences and Engineering Research Council of Canada (NSERC).
\end{acknowledgments}

%%%%%%%%%%%%%%%%%%%%%%%%%%%%%%%%%%%%%%%%%%

\appendix

%%%%%%%%%%%%%%%%%%%%%%%%%%%%%%%%%%%%%%%%%%

\section{Properties of the Kubo susceptibility}\label{Appendix:Kuboproperties}

We study the optical properties of a bulk Chern insulator in terms of the effective dielectric tensor (\ref{effdielectrictensor}). During our discussion of optical activity in Sec.~\ref{Sec:Noveloptics}, we made extensive use of the decomposition of the dielectric tensor into its reactive and absorptive parts, defined in Eqs. (\ref{reactive}) and (\ref{absorptive}). The third term in this dielectric tensor is purely reactive, involving the quantum anomalous Hall response, while the second term, the Kubo susceptibility tensor (\ref{Kubosusc}), has both reactive and absorptive parts that are defined in the same way,
\begin{align}
    \mathscr{R}\big(\chi^{i\ell}(\omega)\big) \equiv \frac{1}{2}\Big(\chi^{i\ell}(\omega) + \big(\chi^{\ell i}(\omega)\big)^*\Big),\nonumber \\
    \mathscr{A}\big(\chi^{i\ell}(\omega)\big) \equiv \frac{1}{2i}\Big(\chi^{i\ell}(\omega) - \big(\chi^{\ell i}(\omega)\big)^*\Big),
\end{align}
such that the full tensor is
\begin{align}
    \chi^{i\ell}(\omega) = \mathscr{R}\big(\chi^{i\ell}(\omega)\big) + i \mathscr{A}\big(\chi^{i\ell}(\omega)\big).
\end{align}
Notably, the Kubo susceptibility can be defined in any number of spatial dimensions, so the following discussion holds in general. When considering time-reversal symmetric systems, these reactive and absorptive parts are just the real and imaginary parts of this tensor. But for time-reversal broken systems this only holds for the diagonal components,
\begin{align}
    \mathscr{R}\big(\chi^{xx}(\omega)\big) &= \Re\big(\chi^{xx}(\omega)\big),\nonumber \\
    \mathscr{A}\big(\chi^{xx}(\omega)\big) &= \Im\big(\chi^{xx}(\omega)\big),
\end{align}
whereas the reactive and absorptive parts of the off-diagonal components of the Kubo susceptibility are generally complex-valued, given by
\begin{align}
    \mathscr{R}\big(\chi^{xy}(\omega)\big) &= \frac{1}{2}\Big(\chi^{xy}(\omega) + \big(\chi^{yx}(\omega)\big)^*\Big),\nonumber \\
    \mathscr{A}\big(\chi^{xy}(\omega)\big) &= \frac{1}{2i}\Big(\chi^{xy}(\omega) - \big(\chi^{yx}(\omega)\big)^*\Big).
\end{align}
When considering time-reversal symmetric systems, the off-diagonal components satisfy $\chi^{xy}(\omega) = \chi^{yx}(\omega)$, and thus these reactive and absorptive parts are just the real and imaginary parts, respectively. But this is no longer true in systems with broken time-reversal symmetry. We assume the  $C_{nz}$ point-group symmetry with $n > 2$ that holds for all known material Chern insulators, where the $z$ direction is orthogonal to the crystal plane where the quantum anomalous Hall effect is observed. From this symmetry, it is straightforward to show that the Kubo susceptibility satisfies
\begin{align}
    \chi^{xx}(\omega) &= \chi^{yy}(\omega),\nonumber \\
    \chi^{xy}(\omega) &= - \chi^{yx}(\omega),
\end{align}
and then the reactive and absorptive components of the off-diagonal components are given by
\begin{align}
    \mathscr{R}\big(\chi^{xy}(\omega)\big) &= i \Im\big(\chi^{xy}(\omega)\big), \nonumber \\
    \mathscr{A}\big(\chi^{xy}(\omega)\big) &= - i \Re\big(\chi^{xy}(\omega)\big),
\end{align}
both of which are purely imaginary. Importantly, the reactive and absorptive parts of the Kubo susceptibility tensor satisfy the Kramers-Kronig relations
\begin{align}
    \mathscr{R}\big(\chi^{i\ell}(\omega)\big) &= \frac{1}{\pi}\, \fint_{-\infty}^{\infty} d\omega'\, \frac{\mathscr{A}\big(\chi^{i\ell}(\omega)\big)}{\omega' - \omega},\nonumber \\
    \mathscr{A}\big(\chi^{i\ell}(\omega)\big) &= - \frac{1}{\pi}\fint_{-\infty}^{\infty} d\omega'\, \frac{\mathscr{R}\big(\chi^{i\ell}(\omega)\big)}{\omega' - \omega},
\end{align}
where the integral notation indicates a Cauchy principal value. These relations guarantee that the response of the medium respects the principle of causality. 

To make sense of these quantities, consider the time-averaged power density that is transferred to the medium by an electromagnetic field, given by
\begin{align}
    \bar{q}(\bm{x}) = \frac{1}{T} \int_{0}^{T} dt\, \bm{J}(\bm{x},t)\cdot\bm{E}(\bm{x},t),
\end{align}
involving the macroscopic current density (\ref{bulkcurrentdensity}). Implementing Fourier-series decompositions of the electric field and the current density,
\begin{align}
    \bm{E}(\bm{x},t) &= \int_{-\infty}^{\infty} \frac{d\omega}{2\pi} \bm{E}(\bm{x},\omega) e^{-i\omega t},\nonumber \\
    \bm{J}(\bm{x},t) &= \int_{-\infty}^{\infty} \frac{d\omega}{2\pi} \bm{J}(\bm{x},\omega) e^{-i\omega t},
\end{align}
the time-averaged power density becomes
\begin{align}
    \bar{q}(\bm{x}) = \frac{1}{2} \int_{0}^{\infty} \frac{d\omega}{2\pi} \Re\big(E_i^*(\bm{x},\omega) J_i(\bm{x},\omega)\big),
\end{align}
which with the constitutive law (\ref{bulkcurrentdensity}) becomes
\begin{align}
    \bar{q}(\bm{x}) = \frac{1}{2} \int_{0}^{\infty} \frac{d\omega}{2\pi} \omega \mathscr{A}\big(\chi^{i\ell}(\omega)\big) E_{i}^*(\bm{x},\omega) E_{\ell}(\bm{x},\omega).
\end{align}
This is a power \textit{density}; the total time-averaged power transferred from the electromagnetic field to the medium is the integrated power density,
\begin{align}
    P_{\mathrm{abs}} = \int d\bm{x}\, \bar{q}(\bm{x}),
\end{align}
which is the electromagnetic power absorbed by the medium. Introducing Fourier modes for the electric field,
\begin{align}
    \bm{E}(\bm{x},\omega) = \sum_{\bm{q}} \bm{E}(\bm{q},\omega) e^{i\bm{q}\cdot\bm{x}},
\end{align}
the total time-averaged power transferred from the electromagnetic field to the medium is
\begin{align}
    P_{\mathrm{abs}} = \frac{1}{2} \int_{0}^{\infty} \frac{d\omega}{2\pi} \omega \mathscr{A}\big(\chi^{i\ell}(\omega)\big) \sum_{\bm{q}} E_{i}^*(\bm{q},\omega) E_{\ell}(\bm{q},\omega).
\end{align}
Therefore, the power that is absorbed by the crystal is controlled by the absorptive part of the Kubo susceptibility tensor, while the part of the Kubo response that does not lead to any absorption is controlled by the reactive part of the Kubo susceptibility tensor, along with the quantum anomalous Hall response described by the third term of Eq. (\ref{effdielectrictensor}). That the absorptive part is related to absorption is especially clear from the explicit forms of these reactive and absorptive parts, given by
\begin{widetext}
\begin{align}
    \mathscr{R}\big(\chi^{i\ell}(\omega)\big) &= e^2 \hbar^2 \sum_{mn} f_{nm} \fint_{\mathrm{BZ}^d} \frac{d\bm{k}}{(2\pi)^d} \frac{1}{(E_{m\bm{k}} - E_{n\bm{k}})^2} \frac{v_{nm}^i(\bm{k}) v_{mn}^{\ell}(\bm{k})}{E_{m\bm{k}} - E_{n\bm{k}} - \hbar\omega},\label{reactKubo} \\
    \mathscr{A}\big(\chi^{i\ell}(\omega)\big) &= \pi e^2 \hbar^2 \sum_{mn} f_{nm} \int_{\mathrm{BZ}^d} \frac{d\bm{k}}{(2\pi)^d}\frac{1}{(E_{m\bm{k}} - E_{n\bm{k}})^2} v_{nm}^i(\bm{k}) v_{mn}^{\ell}(\bm{k})\delta\big(E_{m\bm{k}} - E_{n\bm{k}} - \hbar\omega\big),
    \label{absKubo}
\end{align}
\end{widetext}
The prefactor $f_{nm} = f_n - f_m$ in the absorptive part of the Kubo susceptibility (\ref{absKubo}) is only nonzero if only one of $f_n$ or $f_m$ is nonzero. This means that one of the indices $n,m$ must represent an occupied valence band and the other represents an unoccupied conduction band. Define the quantity $E_{\mathrm{gap}}$ to be the smallest separation between the highest-energy occupied valence band and the lowest-energy unoccupied conduction band across the $\mathrm{BZ}^d$ (the ``bandgap"). If we restrict the frequencies $\omega$ of the Maxwell fields to those for which $\hbar \omega < E_{\mathrm{gap}}$, then the absorptive part of the Kubo susceptibility tensor (\ref{absKubo}) must vanish, even in the absence of time-reversal symmetry.

\section{Bulk transfer matrix algebra}\label{Appendix:Bulktransmissionreflection}

We calculate the bulk transmission and reflection coefficients in Eqs. (\ref{bulktransmission}) and (\ref{bulkreflection}), which are defined by the relations (\ref{bulktensors}). For each of the regions $1 \leq i \leq 3$, we introduce a formal vector quantity
\begin{align}
    \mathcal{E}_i^{\alpha}(z,\omega) \equiv \left(\mqty{E_{i+}^{\alpha}(z,\omega) \\ E_{i-}^{\alpha}(z,\omega)}\right).
    \label{vectorquantity}
\end{align}
We want to relate these quantities at the two interfaces, and we do so by introducing a bulk transfer matrix that is defined implicitly by
\begin{align}
     \mathcal{E}_1^{\alpha}(z = 0,\omega) = \mathcal{M}_{13}^{\alpha\beta}(\omega) \mathcal{E}_3^{\beta}(z = - D, \omega).
     \label{E1E3}
\end{align}
This transfer matrix will involve the Fresnel coefficients for the interfaces $z = 0$ and $z = - D$. We decompose this bulk transfer matrix into a product of three transfer matrices, defined as follows. First we relate the fields at $z = 0$ in region $1$ to those at $z = 0$ in region $2$ by taking
\begin{align}
    \mathcal{E}_{1}^{\alpha}(z = 0,\omega) = M_{12}^{\alpha\beta}(\omega) \mathcal{E}_{2}^{\beta}(z=0,\omega),
    \label{E1E2}
\end{align}
involving a transfer matrix that depends on the Fresnel coefficients for this interface through 
\begin{align}
    M_{12}^{\alpha\beta}(\omega) = \left(\mqty{\big[t_{12}^{-1}(\omega)\big]^{\alpha\beta} & r_{12}^{\alpha\gamma}(\omega)\big[t_{12}^{-1}(\omega)\big]^{\gamma\beta} \\ -\big[t_{12}^{-1}(\omega)\big]^{\alpha\gamma} r_{21}^{\gamma\beta}(\omega) & \big[t_{12}^{-1}(\omega)\big]^{\alpha\beta}}\right),
    \label{M12}
\end{align}
where $r_{21}^{\alpha\beta}(\omega)$ is defined by
\begin{align}
    E_{2-}^{\alpha}(z=0,\omega) = r_{21}^{\alpha\beta}(\omega) E_{2+}^{\beta}(z=0,\omega). 
    \label{inversereflection12}
\end{align}
Next we introduce a kind of ``propagator" that accounts for the propagation of light across the medium, 
\begin{align}
    \mathcal{E}_{2}^{\alpha}(z=0,\omega) = M_2^{\alpha\beta}(D) \mathcal{E}_{2}^{\beta}(z=-D,\omega),
    \label{propagatordef}
\end{align}
which is diagonal for circularly polarized light,
\begin{align}
    M_2^{\alpha\beta}(D) = \left(\mqty{p_2^{\alpha\beta}(D) & 0 \\ 0 & p_{2}^{\alpha\beta}(-D)}\right),
    \label{propagator}
\end{align}
where $p_{2}^{\alpha\beta}(\pm D) = e^{i\tilde{\omega} n_2^{\alpha}(\omega) D} \delta^{\alpha\beta}$. Finally, we relate the fields across the interface $z = - D$ by taking
\begin{align}
    \mathcal{E}_{2}^{\alpha}(z = -D,\omega) = M_{23}^{\alpha\beta}(\omega) \mathcal{E}_{3}^{\beta}(z=-D,\omega),
    \label{E2E3}
\end{align}
involving a transfer matrix that depends on the Fresnel coefficients for this interface through 
\begin{align}
    M_{23}^{\alpha\beta}(\omega) = \left(\mqty{\big[t_{23}^{-1}(\omega)\big]^{\alpha\beta} & r_{23}^{\alpha\gamma}(\omega)\big[t_{23}^{-1}(\omega)\big]^{\gamma\beta} \\ -\big[t_{23}^{-1}(\omega)\big]^{\alpha\gamma} r_{32}^{\gamma\beta}(\omega) & \big[t_{23}^{-1}(\omega)\big]^{\alpha\beta}}\right),
\end{align}
where $r_{32}^{\alpha\beta}(\omega)$ is defined analogously to Eq. (\ref{inversereflection12}). Then from the defining relation (\ref{E1E3}), the bulk transfer matrix is given by
\begin{align}
    \mathcal{M}_{13}^{\alpha\beta}(\omega) = M_{12}^{\alpha\gamma}(\omega) M_{2}^{\gamma\delta}(D)M_{23}^{\delta\beta}(\omega).
    \label{M13pieces}
\end{align}
Multiplying out the explicit forms for the matrices on the right-hand-side given above, and using Eq. (\ref{bulktensors}), the bulk transfer matrix is
\begin{widetext}
\begin{align}
    \mathcal{M}_{13}^{\alpha\beta}(\omega) = \left(\mqty{\mathcal{T}_{31}^{\alpha\beta}(\omega) - \mathcal{R}_{13}^{\alpha\gamma}(\omega) \big[\mathcal{T}_{13}^{-1}(\omega)\big]^{\gamma\delta} \mathcal{R}_{31}^{\delta\beta}(\omega) &  \mathcal{R}_{13}^{\alpha\gamma}(\omega)\big[\mathcal{T}_{13}^{-1}(\omega)\big]^{\gamma\beta} \\ - \big[\mathcal{T}_{13}^{-1}(\omega)\big]^{\alpha\gamma} \mathcal{R}_{31}^{\gamma\beta}(\omega) & \big[\mathcal{T}_{13}^{-1}(\omega)\big]^{\alpha\beta}}\right),
    \label{M13}
\end{align}
from which follows the bulk transmission coefficients
\begin{equation}
    \mathcal{T}_{13}^{\alpha\beta}(\omega) = t_{23}^{\alpha\gamma}(\omega) \bigg[\Big(\mathbb{I} - p_2(D) r_{21}(\omega) p_2(D) r_{23}(\omega)\Big)^{-1}\bigg]^{\gamma\delta} p_2^{\delta\lambda}(D) t_{12}^{\lambda\beta}(\omega),
    \label{T13CI}
\end{equation}
and the bulk reflection coefficients
\begin{equation}
    \mathcal{R}_{13}^{\alpha\beta}(\omega) = r_{12}^{\alpha\beta}(\omega) + t_{21}^{\alpha\gamma}(\omega) p_2^{\gamma\delta}(D) r_{23}^{\delta\lambda}(\omega)\bigg[\Big(\mathbb{I} - p_2(D) r_{21}(\omega) p_2(D) r_{23}(\omega)\Big)^{-1}\bigg]^{\lambda\eta} p_2^{\eta\nu}(D) t_{12}^{\nu\beta}(\omega).
    \label{R13CI}
\end{equation}
\end{widetext}
Working out the matrix inverses in these expressions for light at normal incidence, noting that the only nonzero Fresnel coefficients are given in Eqs. (\ref{Fresnel12}) and (\ref{Fresnel23}), we end up with the bulk coefficients (\ref{bulktransmission}) and (\ref{bulkreflection}).

\section{Sheet transfer matrix algebra}\label{Appendix:Sheettransfermatrix}

We derive the sheet transmission and reflection coefficients in Eqs.~(\ref{Sheettransmissioncoeffs}) and (\ref{Sheetreflectioncoeffs}). For a two-dimensional current sheet at $z=0$, the volumetric current density is given by Eq. (\ref{volumetricJ}). To describe the transmission and reflection of light across this two-dimensional sheet, we solve the Maxwell equations through an analysis based on Green functions with the current density (\ref{sheetcurrentdensity}) as a source. For light propagating along the $z$ direction, the total electric field is
\begin{align}
    \bm{E}(z,\omega) &= \bm{E}_0(z,\omega) + \bm{E}_{s}(z,\omega),
    \label{totalEB}
\end{align}
where the first term is the ``free-space" solution absent the current sheet, while the field sourced by the sheet is
\begin{align}
    \bm{E}_s(z,\omega) &= \frac{i}{\omega} \int_{-\infty}^{\infty} dz'\, \bm{G}_{E}(z - z';\omega)\cdot \bm{J}(z',\omega).
    \label{sourceEB}
\end{align}
involving the electric Green function 
\begin{align}
    \bm{G}_{E}(z - z';\omega) &= 2\pi i \tilde{\omega}  e^{i\tilde{\omega} (z - z')} \Theta(z - z') \hat{\bm{e}}_x\otimes\hat{\bm{e}}_x\nonumber \\
    &+ 2\pi i \tilde{\omega} e^{i\tilde{\omega}(z - z')} \Theta(z - z') \hat{\bm{e}}_y\otimes\hat{\bm{e}}_y\nonumber \\
    &+ 2\pi i \tilde{\omega} e^{-i\tilde{\omega}(z - z')} \Theta(z'- z) \hat{\bm{e}}_x\otimes\hat{\bm{e}}_x\nonumber \\
    &+ 2\pi i \tilde{\omega} e^{-i\tilde{\omega}(z - z')} \Theta(z'- z) \hat{\bm{e}}_y\otimes\hat{\bm{e}}_y\nonumber \\
    &- 4\pi \delta(z - z') \hat{\bm{e}}_z\otimes\hat{\bm{e}}_z,
    \label{electricGTxy}
\end{align}
and $\Theta(z-z')$ is the Heaviside step function. Consider this source field at the position of the sheet $z = 0$. Implementing the electric Green function (\ref{electricGTxy}) in the first line of Eq. (\ref{sourceEB}), and evaluating at $z = 0$, we have
\begin{align}
    \bm{E}_s(z = 0,\omega) = - \frac{2\pi}{c} \bm{K}(\omega),
\end{align}
which we refer to as the \textit{self-field} of the sheet. The sheet current density (\ref{sheetcurrentdensity}) on the right-hand-side is related to the (total) electric field at the sheet. Implementing the solution (\ref{totalEB}) on the right-hand-side, we find
\begin{align}
    \bm{K}(\omega) = \left(\mathbb{I} + \frac{2\pi}{c}\bm{\sigma}_{2d}(\omega)\right)^{-1} \cdot \bm{\sigma}_{2d}(\omega) \cdot \bm{E}_0(\omega).
    \label{sheetcurrentdensity0}
\end{align}
where $\bm{E}_0(\omega) \equiv \bm{E}_0(z=0,\omega)$. Next we consider the source electric field away from the sheet, that is, in the region $z > 0$ above and the region $z < 0$ below the sheet. Implementing the Green function (\ref{electricGTxy}) in the source electric field (\ref{sourceEB}), this \textit{radiated field} is given by
\begin{align}
    \bm{E}_{\mathrm{rad}}(z,\omega) = \bm{E}_{\mathrm{rad}+}(z,\omega) + \bm{E}_{\mathrm{rad}-}(z,\omega),
\end{align}
where the upward and downward propagating radiated fields are
\begin{align}
    \bm{E}_{\mathrm{rad}\pm}(z,\omega) = -\frac{2\pi}{c}e^{\pm i\tilde{\omega}z}\Theta(\pm z)\bm{K}(\omega).
    \label{radiatedsheetfieldappendix}
\end{align}
Here the $+$ field propagates upward into region $1$ and the $-$ field propagates downward away from the sheet. 

We first compute the bare transmission and reflection amplitudes of the isolated sheet when it is embedded in the homogeneous dielectric of region $1$. To distinguish this auxiliary bare-sheet problem from the physical fields in region $3$, we denote the field amplitudes immediately below the sheet in the homogeneous continuation of region $1$ by $\tilde{E}_{1\pm}^{\alpha}(0^-)$. The in-plane symmetry assumed in Sec.~\ref{Sec:Sheetoptics} implies $\sigma_{2d}^{xx}(\omega) = \sigma_{2d}^{yy}(\omega)$ and $\sigma_{2d}^{yx}(\omega) = -\sigma_{2d}^{xy}(\omega)$, so the sheet conductivity is diagonal in the circular basis, and its two eigenvalues are
\begin{align}
    \Sigma^\pm(\omega) \equiv \sigma_{2d}^{xx}(\omega) \pm i\sigma_{2d}^{xy}(\omega).
    \label{Sigmaappendix}
\end{align}
We will henceforth suppress the frequency variable $\omega$ to simplify notation. The signs in $\Sigma^\pm$ label the two eigenvalues of the sheet conductivity, while the subscripts $+$ and $-$ on the fields label their propagation direction. With the circular polarization conventions used in Sec.~\ref{Sec:Sheetoptics}, the downward-propagating components $E_-^L$ and $E_-^R$ sample the conductivity eigenvalues $\Sigma^-$ and $\Sigma^+$, respectively, while the upward-propagating components $E_+^L$ and $E_+^R$ sample the conductivity eigenvalues $\Sigma^+$ and $\Sigma^-$, respectively. Thus, for a downward-propagating field we have
\begin{align}
    K_{-}^{L} &= \Sigma^-\left(E_{1-}^{L}(0^+) - \frac{2\pi}{c}K_{-}^{L}\right), \nonumber \\
    K_{-}^{R} &= \Sigma^+\left(E_{1-}^{R}(0^+) - \frac{2\pi}{c}K_{-}^{R}\right),
    \label{downwardcurrentsappendix}
\end{align}
while for an upward propagating field we have
\begin{align}
    K_{+}^{L} &= \Sigma^+\left(\tilde{E}_{1+}^{L}(0^-) - \frac{2\pi}{c }K_{+}^{L}\right), \nonumber \\
    K_{+}^{R} &= \Sigma^-\left(\tilde{E}_{1+}^{R}(0^-) - \frac{2\pi}{c}K_{+}^{R}\right).
    \label{upwardcurrentsappendix}
\end{align}
Solving these equations gives
\begin{align}
    K_{-}^{L} &= \frac{\Sigma^-}{1 + \frac{2\pi}{c}\Sigma^-}E_{1-}^{L}(0^+), \nonumber \\
    K_{-}^{R} &= \frac{\Sigma^+}{1 + \frac{2\pi}{c}\Sigma^+}E_{1-}^{R}(0^+), \nonumber \\
    K_{+}^{L} &= \frac{\Sigma^+}{1 + \frac{2\pi}{c}\Sigma^+}\tilde{E}_{1+}^{L}(0^-), \nonumber \\
    K_{+}^{R} &= \frac{\Sigma^-}{1 + \frac{2\pi}{c}\Sigma^-}\tilde{E}_{1+}^{R}(0^-).
    \label{barecurrentsappendix}
\end{align}
The reflected field is the radiated field propagating away from the sheet on the side of incidence, while the transmitted field is the incident field continued through the sheet plus the radiated field propagating away from the sheet on the opposite side. Transmission preserves the direction of propagation and therefore preserves the circular label, while reflection reverses the direction of propagation and therefore reverses the circular label. The nonzero bare sheet coefficients are therefore
\begin{align}
    t_{\pm}^{LL} &= \frac{1}{1 + \frac{2\pi}{c}\Sigma^{\pm}}, \qquad
    r_{\pm}^{LR} = -\frac{\frac{2\pi}{c}\Sigma^{\pm}}{1 + \frac{2\pi}{c}\Sigma^{\pm}}, \nonumber \\
    t_{\pm}^{RR} &= \frac{1}{1 + \frac{2\pi}{c}\Sigma^{\mp}}, \qquad
    r_{\pm}^{RL} = -\frac{\frac{2\pi}{c}\Sigma^{\mp}}{1 + \frac{2\pi}{c}\Sigma^{\mp}}.
    \label{baresheetcoefficientsappendix}
\end{align}
These are the bare sheet coefficients: they describe the isolated sheet embedded in the dielectric medium of region $1$, before including the substrate in region $3$.

We dress these bare sheet coefficients by the ordinary interface with the substrate using transfer matrices. The transfer matrix algebra separates into two circular blocks, where the first block contains the downward-propagating $L$ component and the upward-propagating $R$ component. The bare sheet transfer relation in this block is
\begin{align}
    \left(\mqty{E_{1+}^{R}(0^+) \\ E_{1-}^{L}(0^+)}\right) = M^L \left(\mqty{\tilde{E}_{1+}^{R}(0^-) \\ \tilde{E}_{1-}^{L}(0^-)}\right),
    \label{bareLsheettransferrelationappendix}
\end{align}
where the transfer matrix is
\begin{align}
    M^L = \frac{1}{t_{-}^{LL}} \left(\mqty{ t_{-}^{LL}t_{+}^{RR} - r_{2-}^{LR}r_{+}^{RL} & r_{-}^{LR} \\ -r_{+}^{RL} & 1}\right).
    \label{bareLsheetmatrixappendix}
\end{align}
The second block contains the downward-propagating $R$ component and the upward-propagating $L$ component. The bare sheet transfer relation in this block is
\begin{align}
    \left(\mqty{E_{1+}^{L}(0^+) \\ E_{1-}^{R}(0^+)}\right) = M^R \left(\mqty{\tilde{E}_{1+}^{L}(0^-) \\ \tilde{E}_{1-}^{R}(0^-)}\right),
    \label{bareRsheettransferrelationappendix}
\end{align}
where the transfer matrix is
\begin{align}
    M^R = \frac{1}{t_{-}^{RR}} \left(\mqty{t_{-}^{RR}t_{+}^{LL} - r_{-}^{RL}r_{+}^{LR} & r_{-}^{RL} \\ -r_{+}^{LR} & 1}\right).
    \label{bareRsheetmatrixappendix}
\end{align}
Next we consider the ordinary interface between the simple dielectrics in regions $1$ and $3$. Since this interface is not optically active, it is scalar in the circular basis, with
\begin{align}
    t_{13} = \frac{2}{1 + n_3}, \qquad r_{13} = \frac{1 - n_3}{1 + n_3}.
    \label{ordinaryinterfaceappendix}
\end{align}
The corresponding transfer matrix is
\begin{align}
    M_{13} = \frac{1}{t_{13}} \left(\mqty{1 & r_{13} \\ r_{13} & 1}\right),
    \label{ordinaryinterfacetransferappendix}
\end{align}
which is defined so that
\begin{align}
    \left(\mqty{\tilde{E}_{1+}^{R}(0^-) \\ \tilde{E}_{1-}^{L}(0^-)}\right) = M_{13} \left(\mqty{E_{3+}^R(0^-) \\ E_{3-}^L(0^-)}\right),\nonumber \\
    \left(\mqty{\tilde{E}_{1+}^{L}(0^-) \\ \tilde{E}_{1-}^{R}(0^-)}\right) = M_{13} \left(\mqty{E_{3+}^L(0^-) \\ E_{3-}^R(0^-)}\right).
\end{align}
Since the sheet has zero thickness, there is no propagation matrix between the sheet and the ordinary interface. The dressed transfer matrices of the combined sheet-substrate structure are therefore
\begin{align}
    \mathcal{M}_{13}^{L} = M^L M_{13}, \qquad \mathcal{M}_{13}^{R} = M^R M_{13}.
    \label{dressedtransfermatricesappendix}
\end{align}
We first extract the dressed coefficients in the $L$ block, 
\begin{align}
    \left(\mqty{E_{1+}^{R}(0^+) \\ E_{1-}^{L}(0^+)}\right) = \mathcal{M}_{13}^{L}\left(\mqty{E_{3+}^{R}(0^-) \\ E_{3-}^{L}(0^-)}\right),
    \label{Lblocktransferrelationappendix}
\end{align}
so for light incident only from region $1$ ($E_{3+}^{R}(0^-) = 0$), 
\begin{align}
    \mathfrak{t}_{13}^{LL} = \frac{1}{\big[\mathcal{M}_{13}^{L}\big]_{22}},\qquad \mathfrak{r}_{13}^{LR} = \frac{\big[\mathcal{M}_{13}^{L}\big]_{12}}{\big[\mathcal{M}_{13}^{L}\big]_{22}},
    \label{tLLmatrixentryappendix}
\end{align}
where the subscripts outside the square brackets denote the relevant matrix components of the $2\times 2$ matrices $\mathcal{M}_{13}^{L}$. Using Eqs.~(\ref{bareLsheetmatrixappendix}), (\ref{ordinaryinterfacetransferappendix}), and (\ref{dressedtransfermatricesappendix}), the two matrix elements needed here are
\begin{align}
    \big[\mathcal{M}_{13}^{L}\big]_{22} &= \frac{1 - r_{13}r_{+}^{RL}}{t_{-}^{LL}t_{13}},\nonumber \\
    \big[\mathcal{M}_{13}^{L}\big]_{12} &= \frac{r_{-}^{LR}\left(1 - r_{13}r_{+}^{RL}\right) + t_{-}^{LL}r_{13}t_{+}^{RR}}{t_{-}^{LL}t_{13}},
    \label{Lblockmatrixentriesappendix}
\end{align}
and substituting Eq.~(\ref{Lblockmatrixentriesappendix}) into Eq.~(\ref{tLLmatrixentryappendix}) gives
\begin{align}
    \mathfrak{t}_{13}^{LL} &= \frac{t_{-}^{LL}t_{13}}{1 - r_{13}r_{+}^{RL}},\nonumber \\
    \mathfrak{r}_{13}^{LR} &= r_{-}^{LR} + \frac{t_{-}^{LL}r_{13}t_{+}^{RR}}{1 - r_{13}r_{+}^{RL}}.
    \label{Lblockdressedcoefficientsappendix}
\end{align}
Meanwhile, to extract the dressed coefficients in the $R$ block, the transfer relation is
\begin{align}
    \left(\mqty{E_{1+}^{L}(0^+) \\ E_{1-}^{R}(0^+)}\right) = \mathcal{M}_{13}^{R}\left(\mqty{E_{3+}^{L}(0^-) \\ E_{3-}^{R}(0^-)}\right),
    \label{Rblocktransferrelationappendix}
\end{align}
so for light incident only from region $1$ ($E_{3+}^{L}(0^-) = 0$), 
\begin{align}
    \mathfrak{t}_{13}^{RR} = \frac{1}{\big[\mathcal{M}_{13}^{R}\big]_{22}}, \qquad \mathfrak{r}_{13}^{RL} = \frac{\big[\mathcal{M}_{13}^{R}\big]_{12}}{\big[\mathcal{M}_{13}^{R}\big]_{22}},
    \label{Rblockmatrixentryappendix}
\end{align}
where the subscripts outside the square brackets denote the relevant matrix components of the $2\times 2$ matrices $\mathcal{M}_{13}^{R}$. Using Eqs.~(\ref{bareRsheetmatrixappendix}), (\ref{ordinaryinterfacetransferappendix}), and (\ref{dressedtransfermatricesappendix}), the entries are
\begin{align}
    \big[\mathcal{M}_{13}^{R}\big]_{22} &= \frac{1 - r_{13}r_{+}^{LR}}{t_{-}^{RR}t_{13}},\nonumber \\
    \big[\mathcal{M}_{13}^{R}\big]_{12} &= \frac{r_{-}^{RL}\left(1 - r_{13}r_{+}^{LR}\right) + t_{-}^{RR}r_{13}t_{+}^{LL}}{t_{-}^{RR}t_{13}},
    \label{Rblockmatrixentriesappendix}
\end{align}
from which follows
\begin{align}
    \mathfrak{t}_{13}^{RR} &= \frac{t_{-}^{RR}t_{13}}{1 - r_{13}r_{+}^{LR}},\nonumber \\
    \mathfrak{r}_{13}^{RL} &= r_{-}^{RL} + \frac{t_{-}^{RR}r_{13}t_{+}^{LL}}{1 - r_{13}r_{+}^{LR}}.
    \label{Rblockdressedcoefficientsappendix}
\end{align}
Combining the two blocks, the dressed transmission and reflection coefficients are precisely the sheet transmission (\ref{Sheettransmissioncoeffs}) and reflection (\ref{Sheetreflectioncoeffs}) coefficients.

\section{Interface between two Chern insulators}\label{Appendix:InterfaceBBC}

We consider an interface $z = 0$ between two semi-infinite bulk Chern insulators in the regions $1$ ($z > 0$) and $2$ ($z < 0$) with Chern vectors $\bm{C}_{\mathcal{V}_1}$ and $\bm{C}_{\mathcal{V}_2}$. In the quasi-static limit of Sec.~\ref{Sec:Chernmirror}, when both Chern projections $C_{i}^z \equiv \hat{\bm{e}}_z \cdot \bm{C}_{\mathcal{V}_i}$ have the same sign, the moduli of these Fresnel coefficients are found to be
\begin{align}
    \abs{t_{12}^{LL}(0)} &= \abs{t_{12}^{RR}(0)} = \left(\frac{4\abs{C_1^z}}{\abs{C_1^z} + \abs{C_2^z} + 2\sqrt{\abs{C_1^z C_2^z}}}\right)^{1/2}, \nonumber \\
    \abs{r_{12}^{LR}(0)} &= \abs{r_{12}^{RL}(0)} = \left(\frac{\abs{C_1^z} + \abs{C_2^z} - 2\sqrt{\abs{C_1^z C_2^z}}}{\abs{C_1^z} + \abs{C_2^z} + 2\sqrt{\abs{C_1^z C_2^z}}}\right)^{1/2}.
    \label{modFresnelsamesign}
\end{align}
When the Chern projections are equal, $C_{1}^z = C_{2}^z$, the transmission coefficients go to unity and the reflection coefficients go to zero, as expected. Conversely, when the Chern projections have opposite signs, so that either $C_{1}^z > 0$ and $C_{2}^z < 0$ or \textit{vice versa}, the moduli of the Fresnel coefficients are found to be
\begin{align}
    \abs{t_{12}^{LL}(0)} &= \abs{t_{12}^{RR}(0)} = \left(\frac{4\abs{C_1^z}}{\abs{C_1^z} + \abs{C_2^z}}\right)^{1/2}, \nonumber \\
    \abs{r_{12}^{LR}(0)} &= \abs{r_{12}^{RL}(0)} = 1,
    \label{modelFresneloppositesign}
\end{align}
where the second line holds \textit{for any} $\abs{C_1^z}$ and $\abs{C_2^z}$, provided the signs of these Chern projections are opposite.

We calculate the corresponding power transmission and reflection coefficients that are defined by Eq. (\ref{PowerTRdefinition}). Considering the cross products of the circular polarization basis vectors, it is straightforward to show
\begin{align}
    \abs{\hat{\bm{e}}_z \cdot \big\langle \bm{S}_{i\pm}(z = 0,\omega)\big\rangle} = \big\langle S_{i\pm}^{L}(\omega)\big\rangle + \big\langle S_{i\pm}^{R}(\omega)\big\rangle,
\end{align}
where for the incident field
\begin{align}
    \big\langle S_{1-}^{L}(\omega)\big\rangle &= \frac{c}{2\pi} \Re \big(n_{1}^{-}(\omega)\big) \abs{E_{1-}^L(\omega)}^2,\nonumber \\
    \big\langle S_{1-}^{R}(\omega)\big\rangle &= \frac{c}{2\pi} \Re\big(n_{1}^{+}(\omega)\big) \abs{E_{1-}^R(\omega)}^2,
    \label{incidentpowerfluxes}
\end{align}
for the transmitted field
\begin{align}
    \big\langle S_{2-}^{L}(\omega)\big\rangle &= \frac{c}{2\pi} \Re \big(n_{2}^{-}(\omega)\big) \abs{t_{12}^{LL}(\omega)}^2 \abs{E_{1-}^L(\omega)}^2,\nonumber \\
    \big\langle S_{2-}^{R}(\omega)\big\rangle &= \frac{c}{2\pi} \Re \big(n_{2}^{+}(\omega)\big) \abs{t_{12}^{RR}(\omega)}^2 \abs{E_{1-}^R(\omega)}^2,
\end{align}
and for the reflected field
\begin{align}
    \big\langle S_{1+}^{L}(\omega)\big\rangle &= \frac{c}{2\pi} \Re \big(n_{1}^{+}(\omega)\big) \abs{r_{12}^{LR}(\omega)}^2 \abs{E_{1-}^R(\omega)}^2,\nonumber \\
    \big\langle S_{1+}^{R}(\omega)\big\rangle &= \frac{c}{2\pi} \Re \big(n_{1}^{-}(\omega)\big) \abs{r_{12}^{RL}(\omega)}^2 \abs{E_{1-}^L(\omega)}^2.
\end{align}
where we have used the defining relations (\ref{generalFresnelcoeffs}) for the Fresnel coefficients at this interface. It will be useful to define the polarization-resolved power transmission and reflection coefficients
\begin{align}
    T_{12}^{LL}(\omega) &\equiv \frac{\big\langle S_{2-}^{L}(\omega)\big\rangle}{\big\langle S_{1-}^{L}(\omega)\big\rangle} = \frac{\Re(n_{2}^{-}(\omega)\big)}{\Re(n_{1}^{-}(\omega)\big)} \abs{t_{12}^{LL}(\omega)}^2,\nonumber \\
    T_{12}^{RR}(\omega) &\equiv \frac{\big\langle S_{2-}^{R}(\omega)\big\rangle}{\big\langle S_{1-}^{R}(\omega)\big\rangle} = \frac{\Re(n_{2}^{+}(\omega)\big)}{\Re(n_{1}^{+}(\omega)\big)} \abs{t_{12}^{RR}(\omega)}^2,\nonumber \\
    R_{12}^{LR}(\omega) &\equiv \frac{\big\langle S_{1+}^{L}(\omega)\big\rangle}{\big\langle S_{1-}^{R}(\omega)\big\rangle} = \abs{r_{12}^{LR}(\omega)}^2,\nonumber \\
    R_{12}^{RL}(\omega) &\equiv \frac{\big\langle S_{1+}^{R}(\omega)\big\rangle}{\big\langle S_{1-}^{L}(\omega)\big\rangle} = \abs{r_{12}^{RL}(\omega)}^2,
    \label{helicityresolvedTR}
\end{align}
from which follows the (total) power transmission and reflection coefficients
\begin{align}
    T_{12}(\omega) &= \frac{T_{12}^{LL}(\omega) \big\langle S_{1-}^{L}(\omega)\big\rangle + T_{12}^{RR}(\omega) \big\langle S_{1-}^{R}(\omega)\big\rangle}{\big\langle S_{1-}^{L}(\omega)\big\rangle + \big\langle S_{1-}^{R}(\omega)\big\rangle},\nonumber \\
    R_{12}(\omega) &= \frac{R_{12}^{LR}(\omega)\big\langle S_{1-}^{R}(\omega)\big\rangle + R_{12}^{RL}(\omega)\big\langle S_{1-}^{L}(\omega)\big\rangle}{\big\langle S_{1-}^{L}(\omega)\big\rangle + \big\langle S_{1-}^{R}(\omega)\big\rangle}.
    \label{powerTRidentity}
\end{align}

We consider the quasi-static limit in which the leading-order values of the refractive indices for both Chern insulators are
\begin{align}
    n_{i}^{\pm}(\omega) \simeq \sqrt{\frac{2e^2}{\hbar\omega}} \sqrt{\pm C_{i}^{z}},
\end{align}
up to terms of order $O(\hbar\omega/E_{\mathrm{gap}})$. We consider the four possible sign combinations of the Chern projections separately. In the first case where both Chern projections are positive, only the right-handed channel has nonzero incident power, and any left-handed fields are evanescent and carry no static flux across the interface. The time-averaged Poynting power fluxes (\ref{incidentpowerfluxes}) become
\begin{align}
    \big\langle S_{1-}^{L}(\omega)\big\rangle &\simeq 0,\nonumber \\
    \big\langle S_{1-}^{R}(\omega)\big\rangle &\simeq \frac{c}{2\pi} \sqrt{\frac{2e^2}{\hbar\omega}} \sqrt{\abs{C_1^z}} \abs{E_{1-}^R(\omega)}^2.
\end{align}
Using these results in Eq. (\ref{powerTRidentity}) for the power transmission and reflection coefficients, we have
\begin{align}
    T_{12}(\omega) &= T_{12}^{RR}(\omega) \simeq \frac{\Re(n_{2}^{+}(\omega)\big)}{\Re(n_{1}^{+}(\omega)\big)} \abs{t_{12}^{RR}(\omega)}^2,\nonumber \\
    R_{12}(\omega) &= R_{12}^{LR}(\omega) \simeq \abs{r_{12}^{LR}(\omega)}^2,
\end{align}
which together with the Fresnel coefficients (\ref{modFresnelsamesign}) and the helicity-resolved power transmission and reflection coefficients (\ref{helicityresolvedTR}) leads to Eq. (\ref{TRsamesign}). Meanwhile, in the second case where both Chern projections are negative, only the left-handed channel has nonzero incident power, and any right-handed fields are evanescent and carry no flux across the interface. The time-averaged Poynting power fluxes (\ref{incidentpowerfluxes}) become
\begin{align}
    \big\langle S_{1-}^{L}(\omega)\big\rangle &\simeq \frac{c}{2\pi} \sqrt{\frac{2e^2}{\hbar\omega}}\sqrt{\abs{C_1^z}}\abs{E_{1-}^L(\omega)}^2,\nonumber \\ \big\langle S_{1-}^{R}(\omega)\big\rangle &\simeq 0.
\end{align}
Using these results in Eq. (\ref{powerTRidentity}) for the power transmission and reflection coefficients, we have
\begin{align}
    T_{12}(\omega) &= T_{12}^{LL}(\omega) \simeq \frac{\Re(n_{2}^{-}(\omega)\big)}{\Re(n_{1}^{-}(\omega)\big)} \abs{t_{12}^{LL}(\omega)}^2,\nonumber \\
    R_{12}(\omega) &= R_{12}^{RL}(\omega) \simeq \abs{r_{12}^{RL}(\omega)}^2,
\end{align}
which together with the Fresnel coefficients (\ref{modFresnelsamesign}) and the power transmission and reflection coefficients (\ref{helicityresolvedTR}) leads again to Eq. (\ref{TRsamesign}).

We next consider the scenario where the Chern projections have opposite signs, and first the case where $C_{1}^z > 0$ while $C_{2}^z < 0$. The time-averaged Poynting power fluxes (\ref{incidentpowerfluxes}) become
\begin{align}
    \big\langle S_{1-}^{L}(\omega)\big\rangle &\simeq 0,\nonumber \\
    \big\langle S_{1-}^{R}(\omega)\big\rangle &\simeq \frac{c}{2\pi} \sqrt{\frac{2e^2}{\hbar\omega}}\sqrt{\abs{C_{1}^z}} \abs{E_{1-}^R(\omega)}^2.
\end{align}
Using these results in Eq. (\ref{powerTRidentity}) for the power transmission and reflection coefficients, we have
\begin{align}
    T_{12}(\omega) &= T_{12}^{RR}(\omega) \simeq \frac{\Re(n_{2}^{+}(\omega)\big)}{\Re(n_{1}^{+}(\omega)\big)} \abs{t_{12}^{RR}(\omega)}^2,\nonumber \\
    R_{12}(\omega) &= R_{12}^{LR}(\omega) \simeq \abs{r_{12}^{LR}(\omega)}^2,
\end{align}
which with the Fresnel coefficients (\ref{modelFresneloppositesign}) and the power transmission and reflection coefficients (\ref{helicityresolvedTR}) leads to Eq. (\ref{TRoppositesign}). And finally let us consider the other case wherein the Chern projection $C_{1}^z < 0$ and $C_{2}^z > 0$. The time-averaged Poynting power fluxes (\ref{incidentpowerfluxes}) become
\begin{align}
    \big\langle S_{1-}^{L}(\omega)\big\rangle &\simeq \frac{c}{2\pi} \sqrt{\frac{2e^2}{\hbar\omega}}\sqrt{\abs{C_1^z}} \abs{E_{1-}^L(\omega)}^2,\nonumber \\
    \big\langle S_{1-}^{R}(\omega)\big\rangle &\simeq 0.
\end{align}
Using these results in Eq. (\ref{powerTRidentity}) for the power transmission and reflection coefficients, we have
\begin{align}
    T_{12}(\omega) &= T_{12}^{LL}(\omega) \simeq \frac{\Re(n_{2}^{-}(\omega)\big)}{\Re(n_{1}^{-}(\omega)\big)} \abs{t_{12}^{LL}(\omega)}^2,\nonumber \\
    R_{12}(\omega) &\simeq \abs{r_{12}^{RL}(\omega)}^2,
\end{align}
which together with the Fresnel coefficients (\ref{modelFresneloppositesign}) and the power transmission and reflection coefficients (\ref{helicityresolvedTR}) again leads to Eq. (\ref{TRoppositesign}).

%%%%%%%%%%%%%%%%%%%%%%%%%%%%%%%%%%%%%%%%%%

\bibliographystyle{apsrev4-2}
\bibliography{Bibliography.bib}

\end{document}